\documentclass[review,onefignum,onetabnum]{article}
\usepackage[bbgreekl]{mathbbol}
\let\savedbbchi\bbchi
\usepackage{hyperref}
\usepackage[a4paper, total={6in, 9in}]{geometry}
\usepackage{tcolorbox,amsthm}
\usepackage{color} 
\usepackage{footmisc}

\usepackage{graphicx}
\usepackage{color}
\usepackage{amsfonts}
\usepackage{empheq}
\usepackage{amssymb}
\usepackage{amsmath}
\usepackage{bm}
\usepackage{a4wide}
\usepackage{enumitem}
\usepackage{subcaption}
\usepackage{natbib}
\setcitestyle{authoryear} 
\usepackage{authblk}
\newcommand{\ave}[1]{\left\langle#1\right\rangle}

\renewcommand\harvardyearright[1]{.} 

\newtheorem{assumption}{Assumption}[]
\newtheorem{remark}{Remark}[]

\numberwithin{equation}{section}

\let\bbchi\savedbbchi

\newcommand{\cS}{\mathcal{S}}

\newcommand{\Ub}{{\bf U}}
\newcommand{\vb}{{\bf v}}
\newcommand{\hv}{\hat{\bf v}}
\newcommand{\bnu}{\bm \nu}

\newcommand{\x}{{\bf x}}

\newcommand{\hbV}{{\bf \hat{V}}}
\newcommand{\X}{{\bf X}}
\newcommand{\V}{{\bf V}}
\newcommand{\ub}{{\bf u}}

\title{Phenotype-structuring of non-local kinetic models of cell migration driven by environmental sensing}

\author{Tommaso Lorenzi\thanks{Department of Mathematical Sciences ``G. L. Lagrange'', Politecnico di Torino, 10129 Torino, Italy (tommaso.lorenzi@polito.it)}
\and
Nadia Loy\thanks{Department of Mathematical Sciences ``G. L. Lagrange'', Politecnico di Torino, 10129 Torino, Italy (nadia.loy@polito.it)}
\and
Chiara Villa\thanks{Sorbonne Universit\'e, CNRS, Universit\'e de Paris, Inria, Laboratoire Jacques-Louis Lions UMR7598, F-75005 Paris, France. Currently at: Université Paris Cité, CNRS, MAP5, Paris, F-75006, France (chiara.villa@math.cnrs.fr)}
}

\begin{document}
\maketitle
\date{}

\begin{abstract}
The capability of cells to form surface extensions to non-locally probe the surrounding environment plays a key role in cell migration. The existing mathematical models for {migration} of cell populations driven by this non-local form of environmental sensing rely on the simplifying assumption that cells in the population share the same cytoskeletal properties, and thus form surface extensions of the same size. To overcome this simplification, we develop a {kinetic} modelling framework wherein a population of migrating cells is structured by a continuous phenotypic variable that captures variability in structural properties of the cytoskeleton. This framework provides a multiscale representation of {cell migration}, from single-cell dynamics to population-level behaviours, as we start with a microscopic model that describes the dynamics of single cells in terms of stochastic processes. Next, we formally derive the mesoscopic counterpart of this model, which consists of a phenotype-structured kinetic equation {that features a phenotype-dependent non-locality.}
 Then, considering an appropriately rescaled version of this kinetic equation, we formally derive the corresponding macroscopic model, which takes the form of a partial differential equation for the cell number density. {To validate the formal procedures employed to derive the macroscopic model from the microscopic model, through the mesoscopic one}, we first compare the results of numerical simulations of the two models. We then compare numerical solutions of the macroscopic model with the results of cell locomotion assays, to test the ability of the model to recapitulate qualitative features of experimental observations. 
\end{abstract}
 
\section{Introduction}
\subsection*{Biological background}
Cell migration is at the heart of morphogenesis, is essential to ensure successful wound healing, immune response, and tissue homeostasis in adult organisms, and is also central to the progression of different pathologies, such as the metastatic cascade associated with cancer malignancy~\citep{trepat2012cell}. 

A key role in cell migration is played by the capability of cells to form surface extensions, which enable them to probe the surrounding environment even multiple cell diameters away (i.e. to perform non-local environmental sensing)~\citep{friedl2009collective,friedl1998cell,mitchison1996actin}. The size and shape of such surface extensions depend on the cytoskeletal properties of the cells, which have been shown to vary not only between cell populations but also amongst cells within the same population, leading to intra-population heterogeneity in cell locomotive phenotype. For instance, fibroblasts in wound healing have been observed to move following the extension of short or long lamellipodia~\citep{trepat2012cell}; invading tumour cells have been reported to display different morphologies and migratory abilities depending on whether they are more epithelial-like or mesenchymal-like~\citep{jolly2018hybrid}; and it has been demonstrated that myoblasts migrate via the extension of long pseudopods or short leading lamellae~\citep{goodman1989e8}. 

Changes in the cell locomotive phenotype can occur following external stimuli, being these chemical or mechanical. For instance: fibroblasts involved in wound healing may increase their lamellipodia size and speed of migration in response to different growth factors~\citep{trepat2012cell}; epithelial-to-mesenchymal transition in cancer can be induced by both growth factors and increased stiffness of the extracellular matrix (ECM)~\citep{gkretsi2018cell}; and the locomotive phenotype of myoblasts has been shown to depend on the adhesive glycoproteins present on the substrate on which they move, likely due to mechanotransduction~\citep{givant2005laminin,goodman1989e8,miyamoto1998fibronectin,rousselle2020laminin}.  
Independently of the specific pathways responsible for such changes, it is clear that the cytoskeletal structure is not a binary state and this translates into a range of different locomotive phenotypes, which are linked to the characteristics of the surface extensions that are formed by the cells.

In the light of mounting evidence indicating that myoblast cytodifferentiation is deeply affected by the molecular composition of the surrounding ECM, myoblasts locomotion over different ECM components was examined in~\citep{goodman1989e8} by means of \emph{in vitro} stripe assays. In each assay, parallel stripes of two different glycoproteins (i.e. laminin and fibronectin) were absorbed onto the substrates. Myoblasts (i.e. murine myoblasts, cell line MM14dy) were  plated at similar densities at one end of the stripes and were allowed to migrate. Cell migration was then tracked via videomicroscopy. Over laminin stripes, cells migrated generally more rapidly undergoing consecutive cycles of long-pseudopod extension and rapid translocation following release from the substrate. On the other hand, cells moved more slowly on fibronectin stripes (about three times slower than over laminin) and mainly through the formation of short leading lamellae. In \citep{goodman1989e8} it was noted that, along with different motile behaviours, cells adhering to different ECM components displayed different cytoskeletal characteristics. In myoblasts migrating over fibronectin, actin\footnote{Actin is a family of globular multi-functional proteins that form microfilaments in the cytoskeleton.} and $\alpha$-actinin\footnote{$\alpha$-actinin is an actin-binding protein.} quickly organised into stress fibres and vinculin\footnote{Vinculin is a membrane-cytoskeletal protein in focal adhesion plaques, which is involved in linkage of integrin adhesion molecules to the actin cytoskeleton.} organised into focal contacts, while these proteins remain more sparsely distributed within the cytoskeleton of myoblasts migrating over laminin. Whilst the specific pathways through which cell mechanosensing translates into cytoskeletal changes and different locomotive strategies are not discussed in~\citep{goodman1989e8}, the experimental results therein presented highlight the important role that intra-population heterogeneity and environment-induced changes in locomotive phenotype play in cell migration driven by non-local environmental sensing.

\subsection*{Mathematical modelling background and content of the article}
A variety of mathematical models for cell migration driven by non-local environmental sensing have been proposed in the last twenty years -- e.g. see the review~\citep{chen2020mathematical} and references therein. The majority of these models rely on the assumption that cells in the population share the same cytoskeletal properties, which do not evolve in time, and thus sense the surrounding environment by means of surface extensions of the same fixed size. In view of the above biological background, this is clearly a simplification that limits the domain of application of these models and their outputs. 

To overcome such a simplification, building upon previous work on phenotype-structured models of cell movement reviewed in~\citep{lorenzi2024phenotype}, in this article we generalise the non-local kinetic modelling approach of~\citep{loy2020kinetic,loy2020modelling} by developing a modelling framework wherein a population of migrating cells is structured by a continuous phenotypic variable that captures variability in structural properties of the cytoskeleton. Cells with different values of the phenotypic variable form surface projections of different length and can then probe the surrounding environment over regions of different sizes. As a result, in contrast to~\citep{loy2020kinetic,loy2020modelling}, the integration domain for the non-locality in the model considered here is non-constant, as it depends on the phenotypic variable.

The framework provides a multiscale representation of migration of cell populations, from single-cell dynamics to population-level behaviours. In fact, the starting point is a microscopic model that describes the dynamics of single cells in terms of stochastic processes. This model takes into account both cell movement, wherein cell reorientation is driven by non-local sensing of the surrounding environment, and environment-induced changes in the cytoskeletal structure. The mesoscopic counterpart of this model, which is formally derived by extending the limiting procedure employed in~\citep{conte2023non} to the phenotypically heterogeneous scenario considered here, is a phenotype-structured kinetic equation for the distribution of cells in each of phase and phenotype space. Then, considering an appropriately rescaled version of this kinetic equation, we formally derive the corresponding macroscopic model, which takes the form of a partial differential equation (PDE) for the cell number density. Being derived from an underlying model of single-cell dynamics, such a macroscopic equation constitutes a sound population-level model of cell migration. 

To validate the formal procedures employed to derive the macroscopic model from the microscopic model, through the mesoscopic one, we first compare the results of numerical simulations of the two models. We then compare numerical solutions of the macroscopic model with the results of cell locomotion assays from~\citep{goodman1989e8}, to test the ability of the model to recapitulate qualitative features of experimental observations. We finally report on numerical solutions of a reduced macroscopic model, and compare them with numerical simulations of the original model, in order to assess the robustness of the observed patterns of cell migration. The modelling framework presented here provides a mean to investigate the movement of phenotypically heterogeneous cell populations driven by environmental sensing, in scenarios where the extracellular environment may induce changes in direction, speed, and sensing radius of the cells. In particular, in this article we explore how the interplay between environment-induced changes in speed and in sensing radius may impact on the migration of cell populations.

\subsection*{Outline of the article}
In section~\ref{sec:micro}, we formulate the microscopic model, the mesoscopic and macroscopic counterparts of which are then presented in section~\ref{sec:macro}. In section~\ref{sec:numsim}, we report on Monte Carlo simulations of the microscopic model and numerical solutions of the PDE that defines the macroscopic model. In section~\ref{sec:discconcs}, we conclude with a discussion and propose some future research directions.

\section{A microscopic model for the migration of phenotypically heterogeneous cell populations}\label{sec:micro}
We consider a phenotypically heterogeneous cell population, where cells may differ in structural properties of their cytoskeleton which are implicated in cell migration. We focus on the case where the environment surrounding the cells can affect both reorientation processes, which underly cell movement, and mechanisms of phenotypic variation, which drive changes in the cytoskeletal structure.

\subsection{Mathematical modelling of the cell microscopic state}
In order to provide a microscopic-scale description of the dynamics of the cell population, at time $t \in \mathbb{R}^+_0$, where $\mathbb{R}^+_0$ is the set of non-negative real numbers (i.e. $\mathbb{R}^+_0 := \mathbb{R}^+ \cup \{0\}$), we describe the microscopic state of the cells by the quadruplet $(\X_t,V_t,\hbV_t, Y_t)$. The random vector $\X_t \in \mathbb{R}^d$, with $d=1,2,3$ depending on the biological problem under study, represents the spatial position of a cell. Moreover, denoting by $\V_t = V_t \, \hbV_t$ the random vector that represents the cell velocity, $\hbV_t \in \mathbb{S}^{d-1}$ is the unit vector in the direction of $\V_t$ (i.e. the direction of cell polarity and thus of cell motion) and $V_t \in [0,V_{\rm max}]$ is the corresponding speed. Here, $\mathbb{S}^{d-1}$ is the unit sphere boundary in $ \mathbb{R}^d$ and $V_{\rm max} \in \mathbb{R}^+$ is the maximal cell speed. In the remainder of the article, we will also use the compact notation $\mathcal{V}:=[0,V_{\rm max}] \times \mathbb{S}^{d-1}$. Finally, the random variable $Y_t \in [0,1]$ models the cell phenotypic state and captures inter-cellular variability in structural properties of the cytoskeleton. In particular, motivated by the experimental results presented in~\citep{goodman1989e8}, we make the following assumptions: 
\begin{assumption}
\label{ass:mod1}
Smaller values of $Y_t$ correlate with a stiffer cytoskeleton (i.e. a cytoskeleton characterised by a higher level of organisation of actin and $\alpha$-actinin into stress fibres and vininculin into focal contacts), which leads to the formation of shorter cell surface extensions (i.e. lamellae).
\end{assumption}

\begin{assumption}
\label{ass:mod2}
Larger values of $Y_t$ correlate with a more flexible cytoskeleton (i.e. a cytoskeleton characterised by more sparsely distributed actin, $\alpha$-actinin, and vinculin), which leads to the formation of longer cell surface extensions (i.e. pseudopods).
\end{assumption}

\begin{figure}[tbhp]
\centering
\includegraphics[height=0.4\textwidth]{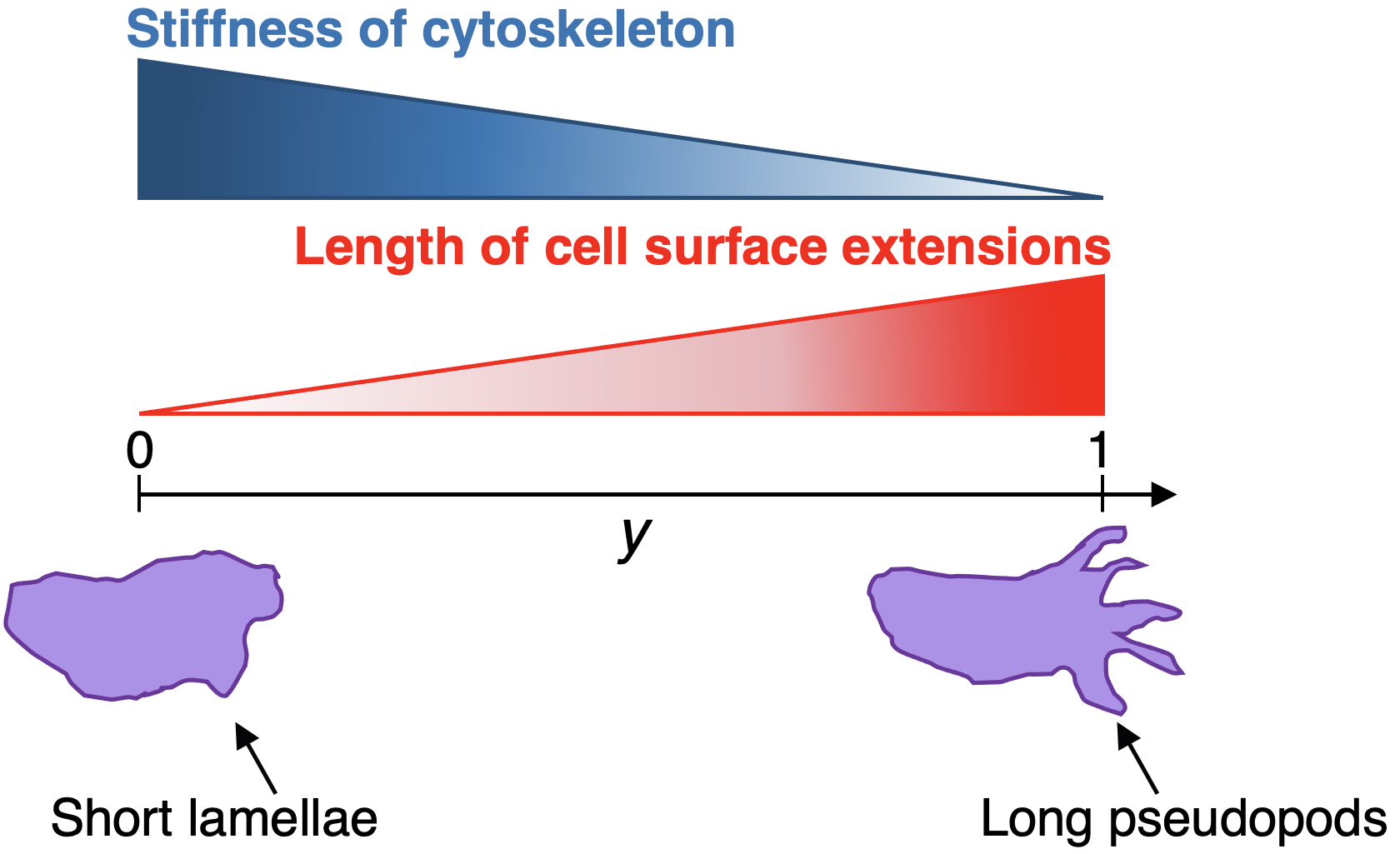}
\caption{Schematics illustrating the relationships between the phenotypic state, $y$, and cell characteristics.  }
\label{Fig1}
\end{figure}

\noindent
These assumptions imply that the movement of cells in phenotypic states close to $0$ is driven by the extension of short lamellae, whereas the movement of cells in phenotypic states close to $1$ is driven by the extension of long pseudopods (cf. the schematics in Figure~\ref{Fig1}). 

The microscopic state of the cells evolves in time due to cell movement, consisting of an alternation of runs over straight lines and reorientations, and phenotypic changes, which we assume to be independent processes. In particular, we model the evolution of the microscopic state $(\X_t,V_t,\hbV_t, Y_t)$ between time $t$ and $t + \Delta t$, with $\Delta t \in \mathbb{R}^+$, through the following system~\citep{conte2023non,pareschi2013interacting}
\begin{equation}\label{eq:micro_stoch}
\begin{cases}
Y_{t+\Delta t}=(1-\Lambda)Y_t+\Lambda Y_t'\\[0.2cm] 
\hbV_{t+\Delta t}=(1-\textrm{M})\hbV_t+\textrm{M} \hbV_t'\\[0.2cm] 
V_{t+\Delta t}=(1-\textrm{M})V_t+\textrm{M} V_t'\\[0.2cm]
\X_{t+\Delta t}=\X_{t}+\Delta t \, V_t \, \hbV_t \, .
\end{cases}
\end{equation}
In the system~\eqref{eq:micro_stoch}, $\Lambda$ and $\textrm{M}$ are independent Bernoulli random variables with parameters $\lambda \Delta t$ and $\mu \Delta t$, respectively, where, as per the modelling strategies introduced in the next subsections,  the parameter $\lambda \in \mathbb{R}^+$ is the rate of phenotypic changes and the parameter $\mu \in \mathbb{R}^+$ is the rate at which cells change their velocity. Note that we are implicitly assuming $\Delta t$ to be small enough so that $\lambda \Delta t \le 1$ and $\mu \Delta t \le 1$. The equations in system~\eqref{eq:micro_stoch} are such that:
\begin{itemize}
\item if $\Lambda=1$ then a phenotypic change occurs and the cell transitions from the original phenotypic state $Y_t$ to a new phenotypic state represented by the random variable $Y_t' \in [0,1]$, whereas if $\Lambda=0$ then no phenotypic change occurs and the cell remains in the phenotypic state $Y_t$; 
\item similarly, if $\textrm{M}=1$ then a velocity change occurs and the cell velocity switches from $(V_t,\hbV_t)$ to $(V_t', \hbV_t') \in \mathcal{V}$, whilst if $\textrm{M}=0$ then no velocity change occurs and the cell keeps its velocity $(V_t,\hbV_t)$;
\item the cell position evolves according to a free-particle drift.
\end{itemize}
We assume $Y'_t$, $\hbV_t'$, and $V_t'$ to be distributed according to the following probability density functions
\begin{equation}\label{eq:micro_vel}
Y_t' \sim K[\mathcal{S}^\ddagger]({y'}|y) \,, \quad \hbV_t' \sim \mathcal{B}[\cS](\hv';y) \,, \quad V_t' \sim \Psi[\cS^\dagger](v';\hv',y) \, .
\end{equation}
As detailed in the next subsections, in~\eqref{eq:micro_vel}:
\begin{itemize}
\item the vector functions 
$$
\mathcal{S}^\ddagger: (t,\x) \in \mathbb{R}^+_0 \times \mathbb{R}^d \mapsto \mathcal{S}^\ddagger(t,\x) \in \mathbb{R}^{m^\ddagger}, \quad \mathcal{S}: (t,\x) \in \mathbb{R}^+_0 \times \mathbb{R}^d \mapsto \mathcal{S}(t,\x) \in \mathbb{R}^{m},
$$
$$
\mathcal{S}^\dagger: (t,\x) \in \mathbb{R}^+_0 \times \mathbb{R}^d \mapsto \mathcal{S}^\dagger(t,\x) \in \mathbb{R}^{m^\dagger},
$$
which are assumed to be given, are introduced in order to incorporate the influence of the surrounding environment on cell dynamics into the model. Specifically: the vector function $\mathcal{S}^\ddagger(t,\x)$ describes the spatial distributions  at time $t$ of $m^\ddagger$ environmental factors that may affect the phenotypic state of the cells; the vector function $\mathcal{S}(t,\x)$ describes the spatial distributions at time $t$ of $m$ environmental factors that may affect the direction of polarity of the cells; and the vector function $\mathcal{S}^\dagger(t,\x)$ describes the spatial distributions at time $t$ of $m^\dagger$ environmental factors that may affect the cell speed. These could be, for instance, the concentrations of abiotic components of the cell microenvironment, such as nutrients, growth factors, adhesion sites on the ECM, and adhesive glycoproteins like laminin and fibronectin.
\item The kernel $K[\mathcal{S}^\ddagger]({y'}|y)$ is a conditional probability density that describes the probability for a cell to transition into the phenotypic state $y'$ given that the cell is in the phenotypic state $y$.
\item The probability density $\mathcal{B}[\cS](\hv';y)$ describes the probability for a cell  in the phenotypic state $y$ to acquire the post-reorientation velocity $\hat{\vb}'$.
\item The probability density $\Psi[\cS^\dagger](v';\hv',y)$ describes the probability for a cell in the phenotypic state $y$ with post-reorientation velocity $\hat{\vb}'$ to acquire the post-reorientation speed $v'$.
\end{itemize}
Note that the probability dentities $K[\mathcal{S}^\ddagger]$, $\mathcal{B}[\cS]$, and $\Psi[\cS^\dagger]$ depend on $(t,\x)$ through the distributions of the environmental factors $\mathcal{S}^\ddagger(t,\x)$, $\cS(t,\x)$, and $\cS^\dagger(t,\x)$.

\subsection{Mathematical modelling of phenotypic changes}
Due to phenotypic changes, which occur at rate $\lambda$, cells at position $\x$ in the phenotypic state $y$ can transition into a new phenotypic state $y'$ with a probability that is given by the kernel $K[\mathcal{S}^\ddagger]({y'}|y)$. The dependence of the phenotypic transition kernel $K$ on the distributions of environmental factors $\mathcal{S}^\ddagger(t,\x)$ captures the fact that phenotypic changes undergone at time $t$ by cells at position $\x$ may be affected by the local cellular microenvironment~\citep{gkretsi2018cell,goodman1989e8,trepat2012cell}. We let the kernel $K\equiv K[\mathcal{S}^\ddagger]$ satisfy the following assumptions
\begin{equation}\label{ass:K0}
K(\cdot |y) >0, \; K(\cdot |y) \in L^1([0,1]), \; \int_0^1 K({y'}|y) \, {\rm d}y' = 1 \;  \text{for  a.e.} \, \, y \in [0,1], \, x\in \Omega, \, t>0,
\end{equation} 
so that $K[\mathcal{S}^\ddagger](\cdot |y) $ is a probability distribution on $[0,1]$.
We then define its average as
\begin{equation}
\label{ass:K1}
y_K[\mathcal{S}^\ddagger](y) :=  \int_0^1 y' \, K[\mathcal{S}^\ddagger]({y'}|y) \, {\rm d}y',
\end{equation}
that is, $y_K[\mathcal{S}^\ddagger](y)$ is the mean phenotypic state of $K[\mathcal{S}^\ddagger](\cdot|y)$ -- i.e. the phenotypic state into which, on average, cells in the phenotypic state $y$ enter as a result of phenotypic changes. This depends on $(t,\x)$ through the distributions of the environmental factors $\mathcal{S}^\ddagger(t,\x)$, and it is such that $y_K: (t,\x,y) \in \mathbb{R}^+_0 \times \mathbb{R}^d \times [0,1]  \to [0,1]$. We also assume that
\begin{equation}\label{hp:K.Linf}
K[\mathcal{S}^\ddagger](y'|\cdot )  \in L^\infty([0,1]) \;  \;  \text{for  a.e.} \, \, y' \in [0,1], \, x\in \Omega, \, t>0.
\end{equation}

\subsection{Mathematical modelling of cell reorientation}
We model cell reorientation as a velocity-jump process~\citep{othmer1988models,stroock1974some}, whereby cells at position $\x$ in the phenotypic state $y$ may change their velocity at rate $\mu$ and acquire a new velocity (i.e. the {\it post-reorientation velocity}), which is prescribed by a turning kernel $T$. In analogy with previous works on the mathematical modelling of cell movement~\citep{chauviere2007modelingb,loy2020kinetic,loy2020modelling}, we assume the cell velocity acquired upon reorientation to be independent from the previous one. Moreover, we let the post-reorientation velocity be affected by cell sensing of the surrounding environment. This is mediated by membrane receptors located along surface extensions, which enable the cells to detect environmental factors within a finite {\it sensing radius} and then adjust the direction of their polarity (i.e. their direction of motion) and their speed accordingly. In particular, here we generalise the modelling strategies presented in~\citep{loy2020kinetic,loy2020modelling} so as to capture cell phenotypic variability by introducing a function 
$$
R : y \in [0,1] \mapsto R(y) \in [R_{\rm min}, R_{\rm max}]  \subset \mathbb{R}^+, \quad 0<R_{\rm min}<R_{\rm max}
$$
that represents the sensing radius of cells in the phenotypic state $y$. Since, under Assumptions~\ref{ass:mod1} and~\ref{ass:mod2}, cells in phenotypic states modelled by larger (smaller) values of $y$ exhibit longer (shorter) surface extensions, and will thus sense environmental factors over a longer (shorter) distance, we assume the function $R$ to be such that
\begin{equation}
\label{ass:Ry}
R(0) = R_{\rm min}, \quad R(1) = R_{\rm max}, \quad \dfrac{{\rm d}}{{\rm d} y} R(y) > 0 \; \forall \, y \in [0,1].
\end{equation}

In this framework, we let the turning kernel $T \equiv T[\mathcal{S},\mathcal{S}^\dagger](\bnu';y)$, which prescribes the post-reorientation velocity $\bnu' = (v',\hv')$ acquired by cells at position $\x$ in the phenotypic state $y$ upon sensing of the distributions of environmental factors $\mathcal{S}$ and $\mathcal{S}^\dagger$ over a neighbourhood of $\x$ of maximal radius $R(y)$ (i.e. the {\it sensing region}), satisfy the following assumptions
\begin{equation}\label{ass:T0}
T(\cdot;y) > 0, \; T(\cdot;y)\in L^1(\mathcal{V}), \; \int_{\mathcal{V}} T(\bnu';y) \, {\rm d}\bnu' = 1  \;  \text{for  a.e.} \, y \in [0,1], \, x \in \Omega, \, t>0,
\end{equation}
so that $T[\mathcal{S},\mathcal{S}^\dagger](\cdot;y)$ is a probability distribution on $\mathcal{V}$. Setting $\vb'=v'\hv'$, we  define
\begin{equation}\label{ass:T1}
\ub_T[\mathcal{S},\mathcal{S}^\dagger](y):= \int_{\mathcal{V}} \vb' \, T[\mathcal{S},\mathcal{S}^\dagger](\bnu';y) \, {\rm d}\bnu' \, ,
\end{equation}
that is, $\ub_T[\mathcal{S},\mathcal{S}^\dagger](y)$ is the mean velocity of $T[\mathcal{S},\mathcal{S}^\dagger](\cdot;y)$ -- i.e. the post-reorientation velocity which is acquired, on average, by cells in the phenotypic state $y$. This depends on $(t,\x)$ through the distributions of the environmental factors $\mathcal{S}(t,\x)$ and $\mathcal{S}^\dagger(t,\x)$, and it is such that $\ub_T : (t,\x,y) \in \mathbb{R}^+_0 \times \mathbb{R}^d \times [0,1] \to \mathcal{V}$.

In particular, building on the modelling strategies presented in~\citep{loy2020kinetic,loy2020modelling}, we use the definition
\begin{equation}\label{def:T_fact}
T[\mathcal{S},\mathcal{S}^\dagger](\bnu';y) := \mathcal{B}[\cS](\hv';y) \, \Psi[\cS^\dagger](v';\hv',y)
\end{equation}
where:
\begin{itemize}
\item The probability density $\mathcal{B}[\cS](\hv';y)$, which prescribes the direction of the post-reorientation velocity $\hat{\vb}'$ of a cell at position $\x$ in the phenotypic state $y$, based on the distributions of environmental factors $\mathcal{S}$ in a neighbourhood of $\x$ of radius $r \in [0,R(y)]$, is defined as
\begin{equation}\label{def:B}
\mathcal{B}[\cS](\hv';y) := \dfrac{\displaystyle{\int_{0}^{R(y)} \gamma_{\mathcal{S}}(r,y) \, b[\mathcal{S}](r,\hv')\,{\rm d}r}}{\displaystyle{\int_{\mathbb{S}^{d-1}}\int_{0}^{R(y)}  \gamma_{\mathcal{S}}(r,y) b[\mathcal{S}](r,\hv')\,{\rm d}r \, {\rm d}\hv'}} \, .
\end{equation}
Details on $\gamma_{\mathcal{S}}(r,y)$ and $b[\mathcal{S}](r,\hv')$ are provided below.

\item The probability density $\Psi[\cS^\dagger](v';\hv',y)$, which prescribes the post-reorientation speed $v'$ of a cell at position $\x$ in the phenotypic state $y$, based on the distributions of environmental factors $\mathcal{S}^\dagger$ in a neighbourhood of $\x$ of radius $r^\dagger \in [0,R(y)]$ along the post-reorientation direction of motion $\hat{\vb}'$, is defined as 
\begin{equation}\label{def:Psi}
\Psi[\cS^\dagger](v';\hv',y) := \dfrac{\displaystyle{\int_{0}^{R(y)} \gamma_{\mathcal{S}^\dagger}(r^\dagger,y) \psi[\mathcal{S}^\dagger(t,\x+r^\dagger \hv')](v')\,{\rm d}r^\dagger}}{\displaystyle{\int_0^{V_\text{max}} \int_{0}^{R(y)} \gamma_{\mathcal{S}^\dagger}(r^\dagger,y) \psi[\mathcal{S}^\dagger(t,\x+r^\dagger \hv')](v')\,{\rm d}r^\dagger\,  {\rm d}v'}} \, .
\end{equation}
Details on $\gamma_{\mathcal{S}^\dagger}(r^\dagger,y)$ and $\psi[\mathcal{S}^\dagger](v')$ are provided below. 
\end{itemize}

In the definitions given by~\eqref{def:B} and~\eqref{def:Psi}:
\begin{itemize}
\item $\gamma_\mathcal{S}: \mathbb{R}^+_0 \times [0,1] \to \mathbb{R}^+_0$ and $\gamma_{\mathcal{S}^\dagger} : \mathbb{R}^+_0 \times [0,1] \to \mathbb{R}^+_0$ are weights  modelling how sensing of the distributions of environmental factors $\mathcal{S}$ and $\mathcal{S}^\dagger$ over neighbourhoods of $\x$ of radii $r \in [0,R(y)]$ and $r^\dagger \in [0,R(y)]$ affects the direction of polarity and the speed of cells in the phenotypic state $y$.  
For instance, if the direction of polarity and the speed of cells in the phenotypic state $y$ are affected only by the environment sensed at the edge of the sensing region, then the functions $\gamma_\mathcal{S}$ and $\gamma_{\mathcal{S}^\dagger}$ can be defined as 
\begin{equation}\label{def:gamma1}
\gamma_\mathcal{S}(r,y) := \delta (r - R(y)) \quad \text{and} \quad \gamma_{\mathcal{S}^\dagger}(r^\dagger,y) := \delta(r^\dagger-R(y)) \,,
\end{equation} 
where $\delta(\cdot)$ is the Dirac delta. Alternatively, these functions may be defined: as Heaviside step functions of $r$ and $r^\dagger$, when the direction of polarity and the speed of the cells are affected by the environmental conditions within the sensing region in a uniform manner; or as monotonically decreasing functions of $r$ and $r^\dagger$, so as to account for a reduced influence of environmental cues closer to the edge of the sensing region~\citep{loy2020kinetic,loy2020modelling}. 

\item $b[\mathcal{S}](r,\hv')$ is a non-negative function that describes how reorientation of cell polarity, driven by the distributions of environmental factors $\mathcal{S}$ in a neighbourhood of $\x$ of radius $r \in [0,R(y)]$, leads the cells to move in direction $\hv'$. For additional details about this function, we refer the reader to~\citep{loy2020kinetic}.
\item $\psi[\mathcal{S}^\dagger(t,\x+r^{\dagger}{\hv}')](v')$ is a probability density that prescribes the post-reorientation speed $v'$, based on the distribution of the environmental factors $\mathcal{S}^\dagger$ in a neighbourhood of $\x$ of radius $r^\dagger \in [0,R(y)]$ along the post-reorientation direction of motion $\hv'$, which is such that
\begin{equation}\label{ass:normpsi}
\int_0^{V_{\rm max}} \psi[\mathcal{S}^\dagger](v') \, {\rm d}v' =1 \,, \quad \int_0^{V_{\rm max}} v' \,  \psi[\mathcal{S}^\dagger](v') \, {\rm d}v' = u_{\psi}[\mathcal{S}^\dagger].
\end{equation}
In~\eqref{ass:normpsi}, $u_{\psi}[\mathcal{S}^\dagger]$ is the mean speed of $\psi[\mathcal{S}^\dagger](v')$, that is, the post-reorientation speed which is acquired, on average, by the cells. This depends on $(t,\x)$ through the distributions of the environmental factors $\mathcal{S}^\dagger(t,\x)$, and it is such that $u_{\psi} : (t,\x) \in \mathbb{R}^+_0 \times \mathbb{R}^d \to [0,V_{\rm max}]$.
\end{itemize}
Overall, the modelling choice corresponding to~\eqref{def:T_fact}-\eqref{def:Psi} implies that at each reorientation there is a simultaneous change of direction and speed: the polarisation direction changes into a new one $\hv'$ and the new speed $v'$ is chosen by sensing the external factor $\cS^\dagger$ along $\hv'$.
Note that here we assume $b[\cS]$ and $\psi[\cS^\dagger]$ to be such 
\[
\mathcal{B}[\cS](\cdot; y) \in L^1(\mathbb{S}^{d-1}), \qquad \Psi[\cS^\dagger](\cdot;\hv',y) \in L^1([0,V_{max}]),
\]
and we also assume $\gamma_\mathcal{S}$ and $\gamma_{\mathcal{S}^\dagger}$ to be such that $\mathcal{B}[\cS]$ and $\Psi[\cS^\dagger]$ are measurable functions of $y$. Under these regularity assumptions, $T[\cS,\cS^\dagger]$ defined via~\eqref{def:T_fact}-\eqref{def:Psi} is a measurable function of $y$. Moreover, using the Fubini-Tonelli theorem, one can see that $T[\cS,\cS^\dagger]$ does satisfy the normalisation assumption in~\eqref{ass:T0}.

\section{Corresponding mesoscopic and macroscopic models}\label{sec:macro}
In this section, we formally derive first the mesoscopic counterpart of the microscopic model presented in the previous section, and then the corresponding macroscopic model.

\subsection{Preliminaries and notation} We let 
\begin{equation}
\label{def:celprobdens}
(\X_t,V_t,\hbV_t, Y_t) \sim f(t,\x,v,\hv,y) \, ,
\end{equation}
where the function $f : \mathbb{R}^+_0 \times \mathbb{R}^d \times [0,V_{\rm max}] \times \mathbb{S}^{d-1} \times [0,1] \to \mathbb{R}^+_0$, which represents the cell distribution in each of phase and phenotype space, is a probability density function, and is thus such that
\begin{equation}
\label{ass:intf1}
\int_{\mathbb{R}^d} \int_{0}^{V_{\rm max}} \int_{\mathbb{S}^{d-1}} \int_{0}^1 f(t,\x,v,\hv,y) \, {\rm d}y \, {\rm d}\hv \, {\rm d}v \, {\rm d}\x = 1 \, .
\end{equation}
In the remainder of the article, we will use the more compact notation
$$
f(t,\x,v,\hv,y) \equiv f(t,\x,\bnu,y), \quad \bnu \equiv (v,\hv) \in \mathcal{V}.
$$
The macroscopic quantities are defined as the statistical moments of $f$ at $(t,\x)$ fixed:
\begin{itemize}
\item the number density (per unit volume) of the cell population (i.e. the cell density)
\begin{equation}\label{def:m}
\rho(t,\x) := \int_{\mathcal{V}} \int_0^1 f(t,\x,\bnu,y) \, {\rm d}y \, {\rm d}\bnu\,;
\end{equation}
\item the normalised distribution of the cell population in the phase space 
\begin{equation}\label{def:p}
p(t,\x,\bnu) := \dfrac{1}{\rho(t,\x)}\int_{0}^1 f(t,\x,\bnu,y)\, {\rm d}y \,;
\end{equation}
\item the mean velocity of the cell population (i.e. the ensemble average velocity)
\begin{equation}\label{def:U}
{\bf U}(t,\x) :=  \int_{\mathcal{V}} \vb \, p(t,\x,\bnu) \,  {\rm d}\bnu
\end{equation}
and the corresponding momentum
\begin{equation}\label{def:rhoU}
\rho(t,\x) \, {\bf U}(t,\x) = \rho(t,\x) \int_{\mathcal{V}} \vb\, p(t,\x,\bnu) \,  {\rm d}\bnu \,;
\end{equation}
\item the second order tensor of the cell population 
\begin{equation}\label{def:rhoD}
\rho(t,\x)\, \mathbb{D}(t,\x) =  \rho(t,\x)\int_{\mathcal{V}} (\vb-{\bf U})\otimes (\vb-{\bf U})  \,p(t,\x,\bnu) \,  {\rm d}\bnu\,,
\end{equation}
being $\mathbb{D}(t,\x)$ the variance-covariance matrix of $p$ for each $(t,\x)$ fixed.
\end{itemize}
Moreover, we will use the following definitions of the counterparts of the above quantities for cells in a given phenotypic state $y$:
\begin{itemize}
\item the normalised distribution of the cell population in the phenotype space
\begin{equation}\label{def:n}
n(t,\x,y) := \dfrac{1}{\rho(t,\x)}\int_{\mathcal{V}} f(t,\x,\bnu,y) \, {\rm d}\bnu \,,
\end{equation}
whereby $\rho(t,\x) n(t,\x,y)$ is the number density of cells in the state $y$; 
\item the mean velocity of cells in the phenotypic state $y$ 
\begin{equation}\label{def:Uy}
{\bf u}(t,\x,y) := \frac{1}{\rho(t,\x)n(t,\x,y)} \int_{\mathcal{V}} \vb  \, f(t,\x,\bnu,y) \,  {\rm d}\bnu,
\end{equation}
and the corresponding momentum
\begin{equation}\label{def:nUy}
\rho(t,\x) n(t,\x,y) \, {\bf u}(t,\x,y) = \int_{\mathcal{V}} \vb \, f(t,\x,\bnu,y) \,  {\rm d}\bnu \,;
\end{equation}
\item the second order tensor of cells in the phenotypic state $y$
\begin{equation}\label{def:nD}
\rho(t,\x)n(t,\x,y)\, \mathbb{d}(t,\x,y) =  \int_{\mathcal{V}} (\vb-{\bf u})\otimes (\vb-{\bf u})  \,f(t,\x,\bnu,y)\,  {\rm d}\bnu\, ,
\end{equation}
where $n(t,\x,y)\mathbb{d}(t,\x,y)$ is the variance-covariance matrix of $f$ with respect to $\bnu$ for each $(t,\x,y)$ fixed.
\end{itemize} 
Finally, we will use the following definition of the mean phenotypic state of the cell population
\begin{equation}\label{eq:ave_phen}
\bar{y}(t,\x) := \int_{0}^1  y \, n(t,\x,y) \, {\rm d}y \, ,
\end{equation}
which corresponds to the average phenotype of the normalised distribution $n$.

\subsection{The corresponding mesoscopic and macroscopic models}\label{mesomacromodels}
The mesoscopic model corresponding to the microscopic model presented in section~\ref{sec:micro} consists of a phenotype-structured kinetic equation for the distribution function $f(t,\x,\bnu,y)$ given in~\eqref{def:celprobdens}, 
the strong form of which reads as
\begin{equation}\label{eq:Boltz_coll_2:strong}
\begin{cases}
&\displaystyle{\partial_t f + \vb \cdot \nabla_\x f = \mu  \left(\rho \, T[\mathcal{S},\mathcal{S}^\dagger] \, n - f \right)} \\
&\displaystyle{\phantom{\partial_t f + \vb \cdot \nabla_\x f =} + \lambda \left(\int_{0}^1 K[\mathcal{S}^\ddagger]({y}|y')  \, f(t,\x,\bnu,y') \,  {\rm d}y' \, - f \right)} \, ,
\\\\
& \displaystyle{n(t,\x,y) := \dfrac{1}{\rho(t,\x)}\int_{\mathcal{V}} f(t,\x,\bnu,y) \, {\rm d}\bnu} \,.
\end{cases}
\end{equation}
This kinetic model is formally derived in Appendix~\ref{appendix} in the Supplementary Materials, starting from system~\eqref{eq:micro_stoch} complemented with relations~\eqref{eq:micro_vel}, using a limiting procedure which generalises the one employed, for instance, in~\citep{conte2023non} to the case of phenotypically heterogeneous cell populations.
\begin{remark}
In~\eqref{eq:Boltz_coll_2:strong}, the term 
$$
\int_{0}^1 K[\mathcal{S}^\ddagger]({y}|y') \, f(t,\x,\bnu,y') {\rm d}y' \, - f
$$
could also be rewritten as a differential term by using the quasi-invariant limit approach for transition probabilities, as similarly done in~\citep{loy2020markov}. 
\end{remark}

Our aim is now to derive a macroscopic model, that is, a set of PDEs describing the evolution of macroscopic quantities which are conserved by the cellular processes (i.e. reorientations and phenotypic changes) incorporated into the mesoscopic model. In order to obtain a closed macroscopic model, different approaches could be adopted in different regimes (see Appendix~\ref{sec:macroclosure} in the Supplementary Materials). Amongst these, in the light of the experimental results presented in~\citep{goodman1989e8} and the considerations made in Appendix~\ref{sec:justscal} in the Supplementary Materials, introducing a small parameter $\varepsilon \in \mathbb{R}^+$, one can consider the following scaling 
\begin{equation}\label{ass:epsscal3}
\lambda = \dfrac{1}{\varepsilon} \, , \quad \mu = \dfrac{1}{\varepsilon} \, ,
\end{equation}
which corresponds to scenarios where phenotypic changes and cell reorientation occur on similar time scales, which are faster than the time scale of spatial dynamics of the cells. The non-dimentionalisation leading to~\eqref{ass:epsscal3} is provided in Appendix~\ref{sec:justscal} in the Supplementary Materials.

 To this end, denoting the solution to the phenotype-structured kinetic equation~\eqref{eq:Boltz_coll_2:strong} under the parameter scaling~\eqref{ass:epsscal3} by $f_{\varepsilon}$, we make the following expansion ansatz (i.e. the Chapman-Enskog expansion)
\begin{equation}\label{ass.CE}
f_{\varepsilon}=f_0+\varepsilon f^{\bot}, \qquad \rho_0:=\int_{\mathcal{V}} \int_0^1 f_0 \, {\rm d}y \,  {\rm d}\bnu= \rho,  \quad \rho^{\bot}:=\int_{\mathcal{V}} \int_0^1 f^{\bot} \, {\rm d}y \, {\rm d}\bnu=0.
\end{equation}
The ansatz~\eqref{ass.CE} implies that the mass is concentrated in the leading order term, $f_0$, while the correction, $f^{\bot}$, carries no mass, and thus
\begin{equation}\label{cond:rho_0}
\rho_\varepsilon=\rho_0 =\rho.
\end{equation}
From~\eqref{ass.CE}, one obtains analogous expansions for the marginals. In fact, integrating $f_\varepsilon/\rho$ over $[0,1]$ and using~\eqref{ass.CE} gives the expansion for the marginal $p_{\varepsilon}$ along with the mass of the corresponding leading order term and of the correction, i.e.
{\small
\begin{equation}\label{def:p_eps}
p_\varepsilon = p_0+\varepsilon p^{\bot}, \quad p_0=\dfrac{1}{\rho}\int_0^1 f_0 \, {\rm d}y, \quad p^{\bot}=\dfrac{1}{\rho}\int_0^1 f^{\bot} \, {\rm d}y, \quad \int_{\mathcal{V}} p_0 \, {\rm d}\bnu =1, \quad \int_{\mathcal{V}}p^{\bot} \, {\rm d}\bnu =0.
\end{equation}
}
Similarly, integrating $f_\varepsilon/\rho$ over $\mathcal{V}$ and using~\eqref{ass.CE} gives the expansion for the marginal $n_{\varepsilon}$ along with the mass of the corresponding leading order term and of the correction, i.e.
{\small
\begin{equation}\label{def:n_eps}
n_\varepsilon =n_0 +\varepsilon n^{\bot}, \quad n_0=\dfrac{1}{\rho}\int_{\mathcal{V}} f_0 \, {\rm d}\bnu, \quad n^{\bot}=\dfrac{1}{\rho}\int_{\mathcal{V}} f^{\bot} \, {\rm d}\bnu, \quad \int_0^1 n_0 \, {\rm d}y=1, \quad \int_0^1 n^{\bot} \, {\rm d}y =0.
\end{equation}
}

\begin{remark}\label{rem:g1}
We now introduce the following integral operator defined on the Banach space $L^1([0,1])$ into itself 
\begin{align*}
&\mathcal{K}: L^1([0,1]) \longrightarrow L^1([0,1])\\
&\mathcal{K}[g] = \int_{0}^1 K[\mathcal{S}^\ddagger]({y}|y') \, g(t,\x,y') \, {\rm d}y',
\end{align*}
which is linear, bounded (cf. assumption~\eqref{hp:K.Linf}), and compact. Because of the positivity assumption in~\eqref{ass:K0}, this operator is strictly positive on the cone of positive functions of $L^1([0,1])$, which is reproducing and also total. Therefore, the Krein-Rutman theorem (see e.g. \cite[Theorem 19.2]{deimling}) allows one to conclude that this integral operator  admits a spectral radius associated with a positive eigenfunction, $g_1(t,\x,y)$, which depends on $(t,\x)$ through $\mathcal{S}^\ddagger(t,\x)$ (i.e. $g_1(t,\x,y) \equiv g_1[\mathcal{S}^\ddagger](y)$) and it is unique up to normalisation, i.e. 
$$
 \int_{0}^1 g_1[\mathcal{S}^\ddagger](y) \, {\rm d}y = 1 \, .
$$ 
In particular, note that, due to the normalisation assumption in~\eqref{ass:K0}, the eigenfunction $g_1(t,\x,y)$ is associated with the eigenvalue $1$.
Moreover, if the phenotypic transition kernel $K[\mathcal{S}^\ddagger]$ also satisfies the following assumption
\begin{equation}\label{ass:Kclosure}
K[\mathcal{S}^\ddagger]({y}|y') \equiv K[\mathcal{S}^\ddagger](y),
\end{equation}
that is, if the new phenotypic states acquired by the cells upon phenotypic changes are independent of the original ones,
then
\[
\mathcal{K}[g] = \int_{0}^1 K[\mathcal{S}^\ddagger]({y}) \, g(t,\x,y') \, {\rm d}y'= K[\mathcal{S}^\ddagger]({y}) \int_{0}^1 g(t,\x,y') \, {\rm d}y'.
\]
In this case, since $g_1$ is the eigenfunction of $\mathcal{K}[g]$ associated with the eigenvalue 1 and it is normalised, then
\begin{equation}\label{eq:g1explicit}
g_1[\mathcal{S}^\ddagger](y) = K[\mathcal{S}^\ddagger](y) \, .
\end{equation}
\end{remark}

In the remainder of this section, we will be letting assumption~\eqref{ass:Kclosure} hold, so that the result~\eqref{eq:g1explicit} holds as well. Moreover, we will be assuming the distributions of environmental factors $\mathcal{S}$, $\mathcal{S}^\dagger$, and $\mathcal{S}^\ddagger$ to be constant in time, i.e. 
\begin{equation}\label{ass:constenvfact}
\mathcal{S}\equiv \mathcal{S}(\x) \, , \quad \mathcal{S}^\dagger\equiv\mathcal{S}^\dagger(\x) \, , \quad  \mathcal{S}^\ddagger \equiv \mathcal{S}^\ddagger(\x),
\end{equation}
in order to consider the scenario where the extracellular environment is stationary. Note that, in this scenario, the result~\eqref{eq:g1explicit} implies that $g_1[\mathcal{S}^\ddagger](y)$ is constant in time.

From this point on, for readability, we simplify the notation by omitting the dependence on $\cS, \cS^\dagger, \cS^\ddagger$, that is,
\[
T[\cS,\cS^\dagger] \equiv T, \qquad T_K[\cS,\cS^\dagger,\cS^\ddagger]\equiv T_K, \qquad \mathcal{T}[\cS,\cS^\dagger,\cS^\ddagger] \equiv \mathcal{T}.
\]

In the framework of assumptions~\eqref{ass:Kclosure} and~\eqref{ass:constenvfact}, exploiting the result~\eqref{eq:g1explicit}, we now employ a limiting procedure that generalises the one used, for instance, in~\citep{hillen2006m}, to the case where a continuous phenotype structure and phenotypic changes are incorporated into the kinetic equation. As similarly done in~\citep{hillen2006m}, we define the operator 
\begin{equation*}
\begin{aligned}[b]
\mathcal{L}^{\bnu}: L^2(\mathcal{V}\times [0,1]) &\longrightarrow L^2(\mathcal{V}\times [0,1])\\[0.1cm]
\varphi & \longmapsto T(\bnu;y)\int_{\mathcal{V}} \varphi(\bnu,y) \, {\rm d}\bnu - \varphi,
\end{aligned}
\end{equation*}
so that $\mathcal{L}^{\bnu}(f) = \rho T(\bnu;y) n \, - f$. Moreover, since, compared to~\citep{hillen2006m}, in our model the microscopic state of the cells  comprises both a velocity component and a phenotype component, the dynamics of which are governed by two independent processes, we introduce the additional operator
\begin{equation*}
\begin{aligned}[b]
\mathcal{L}^y: L^2(\mathcal{V}\times [0,1]) &\longrightarrow L^2(\mathcal{V}\times [0,1])\\
\varphi &\longmapsto K(y)\int_{0}^1 \varphi(\bnu,y) \, {\rm d}y  - \varphi,
\end{aligned}
\end{equation*}
so that $\mathcal{L}^y(f)=\rho \,K(y) \, p \, - f$. Note that, on the same function space $L^2(\mathcal{V}\times [0,1])$, we can also define the sum of the two operators $\mathcal{L} := \mathcal{L}^{\bnu}+\mathcal{L}^y$, with 
\begin{equation}\label{def:L}
\begin{aligned}[b]
\mathcal{L}: L^2(\mathcal{V}\times [0,1]) &\longrightarrow L^2(\mathcal{V}\times [0,1])\\
\varphi &\longmapsto T(\bnu;y)\int_{\mathcal{V}} \varphi(\bnu',y) \, {\rm d}\bnu'+K(y)\int_{0}^1 \varphi(\bnu,y') \, {\rm d}y'  - 2\varphi,
\end{aligned}
\end{equation}
which allows us to rewrite the phenotype-structured kinetic equation~\eqref{eq:Boltz_coll_2:strong} under the parameter scaling~\eqref{ass:epsscal3} as 
\begin{equation}\label{eq:kin_strong_op}
\partial_t f_\varepsilon +\vb\cdot \nabla_\x f_\varepsilon = \dfrac{1}{\varepsilon}\mathcal{L}(f_\varepsilon).
\end{equation}
Substituting~\eqref{ass.CE} into~\eqref{eq:kin_strong_op}, one finds a hierarchy of equations in $\varepsilon$. In $\varepsilon^0$ 
\begin{equation}\label{eq:eps0}
\mathcal{L}(f_0)=0,
\end{equation}
while in $\varepsilon^1$ 
\begin{equation}\label{eq:eps1}
\partial_t f_0 +\vb \cdot \nabla_\x f_0=\mathcal{L}(f^{\bot}).
\end{equation}

The only conserved quantity is the mass (i.e. the number density at $(t,\x)$ fixed), as 
$$
\int_{\mathcal{V}}\int_0^1 \mathcal{L}(f) \, {\rm d} y \, {\rm d} \bnu=0,
$$
while no higher-order moments of $f$ are conserved. As a consequence, in the regime~\eqref{ass:epsscal3}, wherein the operators $\mathcal{L}^{\bnu}$ and $\mathcal{L}^y$ defining the operator $\mathcal{L}$ are of the same order, the macroscopic model will be a closed PDE for the cell density, that is, the number density of the distribution $f_0$ which makes the operator $\mathcal{L}$ vanish. From~\eqref{eq:eps0} one obtains
\[
f_0(t,\x,\bnu,y)=\dfrac{\rho(t,\x)}{2}\left(T(\bnu;y) \, n^0(\x,y)+K(y) \, p^0(\x,\bnu)\right),
\]
which requires the marginal stationary equilibria $n^0(\x,y)$ and $p^0(\x,\bnu)$ that nullify, respectively, the operator corresponding to phenotypic changes and the operator corresponding to cell reorientation. 

Integrating~\eqref{eq:kin_strong_op} with respect to $y$ one recovers the evolution equation for the marginal $p_\varepsilon$, that is,
\begin{equation}\label{eq:p_eps}
\partial_t  \left(\rho p_\varepsilon\right) +\vb \cdot \nabla_\x \left(\rho p_\varepsilon \right) = \dfrac{1}{\varepsilon} \bar{\mathcal{L}}^{\bnu} (f_\varepsilon),
\end{equation}
where 
\begin{equation}\label{def:bLnu}
\begin{aligned}[b]
\bar{\mathcal{L}}^{\bnu}: L^2(\mathcal{V}\times [0,1]) &\longrightarrow L^2(\mathcal{V})\\
\varphi &\longmapsto \int_0^1 T(\bnu;y)\int_{\mathcal{V}} \varphi(\bnu',y) \, {\rm d}\bnu' \, {\rm d}y - \int_0^1 \varphi (\bnu,y) \, {\rm d}y,
\end{aligned}
\end{equation}
while the evolution equation for $n_\varepsilon$ is
\begin{equation}\label{eq:n_eps}
\partial_t (\rho n_\varepsilon) + \nabla_\x \cdot \left(\rho n_\varepsilon \, {\bf u}\right) = \dfrac{1}{\varepsilon} \bar{\mathcal{L}^y}(f_\varepsilon) \, ,
\end{equation}
where
\begin{equation}\label{def:bLy}
\begin{aligned}[b]
\bar{\mathcal{L}^y} : L^2(\mathcal{V}\times[0,1]) &\longrightarrow L^2([0,1])\\
\varphi &\longmapsto K(y)\int_{\mathcal{V}}\int_0^1 \varphi(\bnu,y) \, {\rm d}y \, {\rm d}\bnu - \int_{\mathcal{V}} \varphi(\bnu,y) \, {\rm d}\bnu.
\end{aligned}
\end{equation}

At leading order in $\varepsilon$, from the equations for the marginals~\eqref{eq:p_eps} and \eqref{eq:n_eps} complemented with~\eqref{def:p_eps} and \eqref{def:n_eps}, one finds
\begin{equation}\label{eq:p0.n0}
p_0(\x,\bnu)=  \int_0^1 T(\bnu;y)n_0(\x,y) \, {\rm d}y, \qquad n_0(\x,y)= K(y),
\end{equation}
i.e.
{\small
\begin{equation}\label{eq:p0}
p_0(\x,\bnu)=T_K(\bnu), \quad T_K(\bnu):=\int_0^1 T(\bnu;y) K(y) \, {\rm d}y.
\end{equation}}
Therefore, the distribution $f_0$, i.e. the solution of~\eqref{eq:eps0}, is
\begin{equation}\label{def:eq_0}
f_0(t,\x,\bnu,y)=\rho(t,\x) \, \mathcal{T}(\bnu,y),
\end{equation}
where
\begin{equation}\label{def:eq_0.T}
\mathcal{T}(\bnu,y):= \dfrac{K(y)}{2}\left[T_K(\bnu)+T(\bnu;y) \, \right].
\end{equation}
Note that from~\eqref{def:eq_0.T} one sees that the stationarity assumptions~\eqref{ass:constenvfact} are needed so as to ensure that $\mathcal{T}$ does not depend on the time variable. Note also that $\mathcal{T}$ is a probability distribution on $\mathcal{V}\times [0,1]$, since
\[
\int_{\mathcal{V}}\int_0^1 \mathcal{T}(\bnu,y) \, {\rm d}y \, {\rm d} \bnu =1, \qquad \forall \x \in \mathbb{R}^d,
\]
due to~\eqref{ass:K0}, \eqref{ass:T0}, and to the measurability of $\mathcal{T}$ as a function of $y$, which provides the regularity needed to apply the Fubini-Tonelli theorem and, in so doing, obtain the above normalisation condition. In~\eqref{def:eq_0} we have used~\eqref{cond:rho_0}. We also stress the fact that the time dependence of the distribution $f_0$ given by~\eqref{def:eq_0} is due exclusively to the time dependence of the number density $\rho$, while $p_0$ and $n_0$, which nullify, respectively, the operator $\bar{\mathcal{L}^{\bnu}}$ and the operator $\bar{\mathcal{L}^y}$, do not depend on the time variable, as a result of the stationarity assumptions~\eqref{ass:constenvfact}. Substituting~\eqref{def:eq_0} and \eqref{def:eq_0.T} into~\eqref{eq:eps1} and integrating over $\mathcal{V}\times [0,1]$ gives, at leading order in $\varepsilon$, the PDE
\begin{equation}\label{eq:govern3}
\partial_t \rho + \nabla_\x \cdot \left(\rho \, \Ub_{\mathcal{T}}\right) = 0 \, ,
\end{equation}
where $\Ub_{\mathcal{T}}$ is the average velocity of $\mathcal{T}$ integrated over $[0,1]$, which is defined as
\begin{equation}\label{def:UT}
\Ub_{\mathcal{T}}  := \int_0^1 \int_{\mathcal{V}} \vb \, \mathcal{T}(\bnu,y) \, {\rm d}\bnu \, {\rm d}y .
\end{equation}
Exploiting the regularity of the phenotypic transition kernel $K$ and the turning kernel $T$, it is possible to invert the order of integration, and, in so doing, obtain
\begin{equation}\label{def:UTKeps0red}
\Ub_{\mathcal{T}}  = \int_{0}^1 \ub_T(y) \,  K(y) \, {\rm d}y,
\end{equation}
with $\ub_T$ being defined via~\eqref{ass:T1}. 

\subsection{Determination of the first order correction to~\eqref{eq:govern3}}\label{sec:correction}
We now seek the first order correction in $\varepsilon$ for~\eqref{eq:govern3}. 
Starting from the kinetic level, we need to consider~\eqref{eq:kin_strong_op} alongside~\eqref{ass.CE}. Including only the terms up to order 1 in $\varepsilon$ yields
\begin{equation}\label{eq:eps12}
\partial_t f_0 +\vb \cdot \nabla_\x\left( f_0+\varepsilon f^{\bot}\right)=\dfrac{1}{\varepsilon}\mathcal{L}(f_0)+\mathcal{L}(f^{\bot}),
\end{equation}
where the first term on the right-hand side vanishes uniformly in $\varepsilon$ due to~\eqref{eq:eps0}. 
To this purpose, we need to determine the correction $f^{\bot}$ to the distribution $f_0$, as $f^{\bot}$ carries the information about the time evolving mass in the transient. 
To do so, we will need to invert the operators $\mathcal{L}, \bar{\mathcal{L}}^{\bnu}$ and $\bar{\mathcal{L}}^{y}$ and to define a Fredholm alternative. 
\subsubsection{Inversion of the operators}
On $L^2(\mathcal{V}\times [0,1])$ we define the inner product 
\begin{equation}\label{def:sc.pr}
h,g  \in L^2(\mathcal{V}\times [0,1]) \quad  \langle h,g \rangle :=  \int_{\mathcal{V}}\int_0^1 h(\bnu,y) g(\bnu,y) \mathcal{T}^{-1}(\bnu,y) \, {\rm d}y \, {\rm d}\bnu,
\end{equation}
so that $L^2(\mathcal{V}\times [0,1])$ endowed with this metric is a Hilbert space.
As the kernel of $\mathcal{L}$ is
\[
\ker(\mathcal{L})= \textrm{span}\lbrace\mathcal{T}\rbrace ,
\]
the operator $\mathcal{L}$ will be inverted on the orthogonal to its kernel $\textrm{span}\lbrace\mathcal{T}\rbrace^{\bot}$, that is, the subspace of functions which are orthogonal to $\mathcal{T}$ according to the scalar product defined via~\eqref{def:sc.pr}. Such an orthogonality condition implies the fact that the functions $\varphi=\varphi(\bnu,y)$ in the subspace $\textrm{span}\lbrace\mathcal{T}\rbrace^{\bot}$ have zero number density, i.e. their integral over $\mathcal{V}\times [0,1]$ is zero. However, the gain term of the operator $\mathcal{L}$ does not feature the density of $\varphi$, but only the density of its marginals, namely $\int_{\mathcal{V}} \varphi(\bnu',y) \, {\rm d}\bnu'$ and $\int_{0}^1 \varphi(\bnu,y') \, {\rm d}y'$, which do not vanish in general, even when $\varphi \in \textrm{span}\lbrace\mathcal{T}\rbrace^{\bot}$. Therefore, the pseudo-inverse is defined as 
\begin{equation}\label{def:L.inv}
\begin{split}
\mathcal{L}^{-1}: \textrm{span}\lbrace\mathcal{T}\rbrace^{\bot} &\longrightarrow \textrm{span}\lbrace\mathcal{T}\rbrace^{\bot}\\
\eta &\longmapsto \varphi \;\; \\
\text{with $\varphi$ s.t.} \;\; \varphi(\bnu,y) =\dfrac{1}{2}&\left[-\eta(\bnu,y)+T(\bnu;y)\int_{\mathcal{V}} \varphi(\bnu',y) \, {\rm d}\bnu'+K(y)\int_{0}^1 \varphi(\bnu,y') \, {\rm d}y'\right].
\end{split}
\end{equation}
We also need to invert the operators $\bar{\mathcal{L}}^{\bnu}$ and $\bar{\mathcal{L}}^y$ defined via~\eqref{def:bLnu} and \eqref{def:bLy}. To this aim, we define the inner products
\begin{equation*}
h,g  \in L^2(\mathcal{V}) \quad  \langle h,g \rangle :=  \int_{\mathcal{V}} h(\bnu) g(\bnu) T_K^{-1}(\bnu) \,  {\rm d}\bnu,
\end{equation*}
and
\begin{equation*}
h,g  \in L^2( [0,1]) \quad  \langle h,g \rangle :=  \int_0^1 h(y) g(y) K^{-1}(y) \, {\rm d}y.
\end{equation*}
Let us now consider the kernels of the operators, that are
$$
\ker (\bar{\mathcal{L}}^{\bnu})= \left\{ \varphi \in L^2(\mathcal{V}\times [0,1]) : \int_{0}^1 \varphi(\bnu,y) \, {\rm d}y = \int_{\mathcal{V}}\int_0^1 T(\bnu;y) \varphi(\bnu',y) \, {\rm d}y \, {\rm d}\bnu' \right\},
$$
and
\begin{equation*}
\begin{split}
\ker (\bar{\mathcal{L}}^y)&= \left\{ \varphi \in L^2(\mathcal{V}\times [0,1]) : \int_{\mathcal{V}} \varphi(\bnu,y) \, {\rm d}\bnu =K(y) \int_{\mathcal{V}}\int_0^1 \varphi(\bnu',y') \, {\rm d}y' \, {\rm d}\bnu' \right\}\\
 &= \textrm{span} \lbrace K \rbrace.
\end{split}
\end{equation*}
In particular, it is worth noting that $\ker (\bar{\mathcal{L}}^{\bnu})$ is not $\textrm{span}\lbrace T_K \rbrace$, unless $ T_K(\bnu) =\int_0^1 T(\bnu;y ) \, {\rm d} y$, which corresponds, for instance, to a uniform kernel $K$ or to the case in which $T$ does not depend on $y$. As a consequence, we have that
\begin{equation}\label{kernel.inclusion}
\ker(\mathcal{L}) \subset \ker (\bar{\mathcal{L}}^{\bnu}) \cap \ker (\bar{\mathcal{L}}^y),
\end{equation}
which is not an identity in general, as we would need $\ker (\bar{\mathcal{L}}^{\bnu}) = \textrm{span}\lbrace T_K \rbrace$. This is the underlying reason for the impossibility of inverting $\mathcal{L}$ in a classical fashion, which is tied to the non-vanishing gain term in $\mathcal{L}$ when it is computed on an element of $\textrm{span}\lbrace\mathcal{T}\rbrace^{\bot}$.
The same issue occurs when inverting $\bar{\mathcal{L}}^{\bnu}$ on $\textrm{span}\lbrace\mathcal{T}\rbrace^{\bot}$. In fact, due to the dependence of $T$ on $y$, the density $\int_\mathcal{V}\int_0^1 \varphi(\bnu',y') \,{\rm d} y' \, {\rm d}\bnu'$ does not appear in the gain term of $\bar{\mathcal{L}}^{\bnu}$. The operator $\bar{\mathcal{L}}^{\bnu}$ should be inverted on the orthogonal to its kernel, which in general is not $\textrm{span}\lbrace T_K \rbrace$. However, as we shall need to invert $\mathcal{L}$, $\bar{\mathcal{L}}^{\bnu}$, and $\bar{\mathcal{L}}^{y}$ at the same time, it is worth noticing that if $\varphi \in \ker(\mathcal{L})$ then, in view of~\eqref{kernel.inclusion}, $\varphi \in \ker (\bar{\mathcal{L}}^{\bnu}) \cap \ker (\bar{\mathcal{L}}^y)$. Since $\varphi \in \ker (\bar{\mathcal{L}}^{\bnu})$, then
\begin{equation*}
\begin{split}
\int_0^1 \varphi(\bnu,y) \, {\rm d} y &= \int_{\mathcal{V}\times[0,1]} \varphi(\bnu',y) T(\bnu;y) \,{\rm d}y \, {\rm d}\bnu' \\
&=\int_{[0,1]} \varphi(\bnu',y') \, {\rm d} y' \int_{\mathcal{V}\times [0,1]} T(\bnu;y)K(y) \, {\rm d} y {\rm d} \bnu'\\
&=  \int_{\mathcal{V}\times[0,1]} \varphi(\bnu',y') \, {\rm d} y' \, {\rm d} \bnu'  T_K(\bnu),
\end{split}
\end{equation*}
where the second equality follows from the fact that we also have $\varphi \in \ker (\bar{\mathcal{L}}^y)$.
Hence, the pseudo-inverse functional $\bar{\mathcal{L}}^{{\bnu}^{-1}}$ is defined on $\textrm{span}\lbrace T_K \rbrace^{\bot}$ and reads
\begin{equation}\label{def:Lnu.inv}
\begin{aligned}[b]
\bar{\mathcal{L}}^{{\bnu}^{-1}}: \textrm{span}\lbrace T_K \rbrace^{\bot} &\to \textrm{span}\lbrace\mathcal{T}\rbrace^{\bot}\\
\eta &\longmapsto  \varphi\\
\;\; \text{with $\varphi$ s.t.} \;\;  &\int_0^1 \int_{\mathcal{V}} T(\bnu;y) \varphi(\bnu',y) \, {\rm d}y \, {\rm d}\bnu' - \int_{0}^1 \varphi(\bnu,y) \, {\rm d}y=\eta(\bnu).
\end{aligned}
\end{equation}
On the other hand, when inverting $\bar{\mathcal{L}}^y$ on $\varphi \in \textrm{span}\lbrace\mathcal{T}\rbrace^{\bot}$, then the gain term of  $\bar{\mathcal{L}}^y$ vanishes. Therefore, the pseudo-inverse reads
\begin{equation}\label{def:Ly.inv}
\begin{aligned}[b]
\bar{\mathcal{L}}^{y^{-1}}: \textrm{span}\lbrace K \rbrace^{\bot} &\to \textrm{span}\lbrace\mathcal{T}\rbrace^{\bot}\\
\eta &\longmapsto  \varphi \;\; \text{with $\varphi$ s.t.} \;\;  -\int_{\mathcal{V}} \varphi(\bnu,y) \, {\rm d} \bnu=\eta(y).
\end{aligned}
\end{equation}

\subsubsection{Computation of the correction}
Going back to the problem of determining $f^{\bot}$ in~\eqref{eq:eps12}, we remark that since $\rho^{\bot}=0$ because of~\eqref{ass.CE}, so that $f^{\bot} \in \textrm{span}\lbrace \mathcal{T} \rbrace^{\bot}$, and thanks to~\eqref{eq:govern3}, one can apply the pseudo-inverse $\mathcal{L}^{-1}$~\eqref{def:L.inv} to the leading order equation~\eqref{eq:eps1} in order to determine $f^{\bot}$. In so doing, one obtains
\begin{equation*}
f^{\bot}=-\dfrac{1}{2}(\partial_t f_0+\vb \cdot \nabla_\x f_0)+\dfrac{1}{2}\left(\rho p^{\bot}K+\rho n^{\bot} T\right),
\end{equation*}
from which, using~\eqref{def:eq_0} along with the fact that $\mathcal{T}$ does not depend on the time variable, which allows one to state that $\partial_t f_0=\mathcal{T}\partial_t \rho$, and the PDE~\eqref{eq:govern3} for expressing $\partial_t \rho =-\nabla_\x\cdot (\Ub_{\mathcal{T}} \rho)$, one finds
\begin{equation}\label{eq:fbot}
f^{\bot}=\dfrac{1}{2}(\nabla_\x\cdot (\Ub_{\mathcal{T}} \rho) \mathcal{T}-\vb \cdot \nabla_\x (\rho \mathcal{T}))+\dfrac{1}{2}\left(\rho p^{\bot} K+\rho n^{\bot} T\right).
\end{equation}
An analogous inversion procedure can be applied to both~\eqref{eq:p_eps} and~\eqref{eq:n_eps} so as to determine $p^{\bot}$ and $n^{\bot}$. One has
\begin{equation}\label{eq:pbot}
\partial_t (\rho p_0) +\vb \cdot \nabla_\x \rho p_0= \bar{\mathcal{L}^{\bnu}}(f^{\bot}),    
\end{equation}
 and
\begin{equation}\label{eq:nbot}
\partial_t (\rho n_0) + \nabla_\x \cdot \left[\rho n_0 \, \dfrac{1}{2} \, \left({\bf u}_T + {\bf U}_{\mathcal{T}} \right)\right]=\bar{\mathcal{L}}^y (f^{\bot}), 
\end{equation}
where, recalling that $f^{\bot} \in \textrm{span}\lbrace \mathcal{T} \rbrace^{\bot}$, we have $\bar{\mathcal{L}}^{\bnu} (f^{\bot}) =\rho \int_0^1 n^{\bot}T(\bnu;y) \, {\rm d} y-\rho p^{\bot}$ and $\bar{\mathcal{L}}^{y} (f^{\bot}) = -\rho n^{\bot}$.
Using the definition~\eqref{eq:p0.n0}-\eqref{eq:p0} of $p_0$, and the fact that integrating the left-hand side of~\eqref{eq:pbot} over $\mathcal{V}$ gives~\eqref{eq:govern3}, one can apply the pseudo-inverse $\bar{\mathcal{L}}^{\bnu^{-1}}$ defined in~\eqref{def:Lnu.inv} and obtain
\[
\rho p^{\bot}=\nabla_\x \cdot (\Ub_{\mathcal{T}} \rho) T_K-\vb \cdot \nabla_\x (\rho T_K) + \rho \int_0^1 n^{\bot}T(\bnu;y) \, {\rm d} y. 
\]
Using the definition~\eqref{eq:p0.n0} of $n_0$, and the fact that integrating the left-hand side of~\eqref{eq:nbot} over $[0,1]$ gives~\eqref{eq:govern3}, one can apply the pseudo-inverse $\bar{\mathcal{L}^{y}}^{-1}$ defined in~\eqref{def:Ly.inv} and obtain
\[
\rho n^{\bot}=\nabla_\x\cdot(\Ub_{\mathcal{T}}\rho) K-\nabla_\x \cdot \left[\dfrac{\rho}{2} \, \left({\bf u}_T + {\bf U}_{\mathcal{T}} \right) K \right].
\]
It can be easily verified that $n^{\bot}$ verifies the last property in~\eqref{def:n_eps}, so that $n^{\bot} \in \textrm{span}\lbrace K \rbrace^{\bot}$. 
As a consequence
\begin{equation*}
\begin{split}
\rho p^{\bot}=&\nabla_\x \cdot (\Ub_{\mathcal{T}} \rho) T_K-\vb \cdot \nabla_\x (\rho T_K) \\
+& \nabla_\x\cdot(\Ub_{\mathcal{T}}\rho)\int_0^1  K(y)T(\bnu;y) \, {\rm d} y-\int_0^1\nabla_\x \cdot \left[\dfrac{\rho}{2} \, \left({\bf u}_T(y) + {\bf U}_{\mathcal{T}} \right) K (y)\right]T(\bnu;y) \, {\rm d} y.
\end{split}
\end{equation*}
It can easily be verified that $p^{\bot}$ satisfies the last property in~\eqref{def:p_eps}, and thus $p^{\bot} \in \textrm{span}\lbrace T_K \rbrace^{\bot}$. In both cases, we have exploited the fact that $p_0$ and $n_0$ do not depend on the time variable, and we have used the PDE \eqref{eq:govern3} for expressing the time derivative $\partial_t\rho$.
Substituting the above expressions for $\rho p^{\bot}$ and $\rho n^{\bot}$ into~\eqref{eq:fbot}, using~\eqref{def:eq_0} and \eqref{def:eq_0.T}, and rearranging terms one obtains 
{\small
\begin{equation}\label{eq:fbot_det}
\begin{aligned}[b]
f^{\bot}&= -\left[\mathcal{T}(\vb-\Ub_{\mathcal{T}})\cdot \nabla_\x\rho+ \left(\vb\cdot \nabla_\x \mathcal{T}-\mathcal{T}\nabla_\x\cdot \Ub_{\mathcal{T}}\right)\rho\right]\\
&\phantom{=}+\dfrac{1}{2}\left[\nabla_\x \cdot \left(\rho\Ub_{\mathcal{T}} \right) \mathcal{T}-\vb \cdot \nabla_\x \left(\rho \mathcal{T}\right)\right]\\
&\phantom{=}+ \dfrac{1}{2}\left[\vb \rho (T_K+T)\cdot \nabla_\x K +\vb \cdot \nabla_\x(\rho T)K -\nabla_\x \cdot \left(\dfrac{\rho}{2} \, \left({\bf u}_T + {\bf U}_{\mathcal{T}} \right) K \right) T\right]\\
& \phantom{=}+\dfrac{1}{2}\left[\int_0^1 \nabla_\x\cdot (\rho\Ub_{\mathcal{T}}) K(y) T(\bnu;y) \, {\rm d}y-\int_0^1\nabla_\x \cdot \left[\dfrac{\rho}{2} \, \left({\bf u}_T(y) + {\bf U}_{\mathcal{T}} \right) K(y)\right]T(\bnu;y) \, {\rm d} y\right]K.
\end{aligned}
\end{equation} 
}
We now introduce the variance-covariance matrix of $\mathcal{T}$, i.e.
{\small
\begin{equation*}
\mathbb{D}_{\mathcal{T}} :=\int_{\mathcal{V}} \int_{0}^1 (\vb-{\bf U}_{\mathcal{T}})\otimes (\vb-{\bf U}_{\mathcal{T}})  \,{\mathcal{T}}(\bnu,y)  \, {\rm d}y \, {\rm d}\bnu.
\end{equation*}}
Exploiting the regularity of $K$ and $T$ to invert the order of integration over $\mathcal{V}\times [0,1]$, one sees that
\begin{equation}\label{def:DT}
\mathbb{D}_{\mathcal{T}}=\int_{\mathcal{V}} \int_{0}^1 (\vb-{\bf U}_{\mathcal{T}})\otimes (\vb-{\bf U}_{\mathcal{T}})  \,T(\bnu;y)  K(y) \, {\rm d}y \, {\rm d}\bnu. 
\end{equation}
Noting also that $\displaystyle{\mathbb{D}_{\mathcal{T}}=\int_{\mathcal{V}} \int_{0}^1 \vb\otimes (\vb-{\bf U}_{\mathcal{T}})  \,{\mathcal{T}}(\bnu,y)  \, {\rm d}y \, {\rm d}\bnu}$ thanks to~\eqref{def:UT},
substituting~\eqref{eq:fbot_det} into~\eqref{eq:eps12} and integrating over $\mathcal{V}\times[0,1]$ yields
\begin{equation}\label{eq:govern3_corr_0}
\begin{split}
\partial_t \rho + &\nabla_\x \cdot \left[\rho \, \Ub_{\mathcal{T}} \left(1 - \varepsilon \, \nabla_\x \cdot \Ub_{\mathcal{T}}\right)\right] = \varepsilon \, \nabla_\x \cdot \nabla_\x\cdot(\mathbb{D}_{\mathcal{T}} \rho)\\
&\phantom{\nabla_\x \cdot [\rho \, \Ub_{\mathcal{T}} (1 }+\dfrac{\varepsilon}{2}\nabla_\x\cdot \left[\left(\mathbb{C}_T-\Ub_{\mathcal{T}}\otimes\Ub_\mathcal{T}\right)\nabla_\x \rho + \left(\boldsymbol{c}_T-\Ub_{\mathcal{T}}\nabla_\x \cdot\Ub_{\mathcal{T}}\right)\rho\right]  \, ,
\end{split}
\end{equation}
where $\Ub_{\mathcal{T}}$ and $\mathbb{D}_{\mathcal{T}}$ are defined via~\eqref{def:UTKeps0red} and \eqref{def:DT}, while
\begin{equation}\label{def:CT}
\mathbb{C}_T := \int_0^1 \ub_T(y) \otimes \ub_T(y) K(y) \, {\rm d} y, \qquad \boldsymbol{c}_T:= \int_0^1 \nabla_\x \cdot (\ub_T(y)) \ub_T(y) K(y) \, {\rm d} y.
\end{equation}
The order $\varepsilon$ drift and diffusion terms in~\eqref{eq:govern3_corr_0} come from the term $\int_{\mathcal{V}}\int_0^1 \vb f^{\bot} \, {\rm d} y {\rm d} \bnu$, and, specifically, from the first line on the right-hand side of~\eqref{eq:fbot_det} through steps analogous to the ones carried out, for instance, in~\citep{hillen2006m}, and they are the usual correction terms related to the collective equilibrium $\mathcal{T}$. Conversely, the second, third, and fourth lines on the right-hand side of~\eqref{eq:fbot_det} lead to the order $\varepsilon/2$ new correction terms, which are due to phenotypic heterogeneity.
We remark that when the kernel $K$ is a Dirac delta or when $\ub_T$ does not depend on the phenotypic variable, then the $\varepsilon/2$-correction terms vanish. Therefore, in these cases equation~\eqref{eq:govern3_corr_0} reduces to
\begin{equation}\label{eq:govern3_corr}
\partial_t \rho + \nabla_\x \cdot \left[\rho \, \Ub_{\mathcal{T}} \left(1 - \varepsilon \, \nabla_\x \cdot \Ub_{\mathcal{T}} \right)\right]= \varepsilon \, \nabla_\x \cdot \nabla_\x\cdot(\mathbb{D}_{\mathcal{T}} \rho)  \, .
\end{equation}
Note that, thanks to~\eqref{cond:rho_0}, the PDE~\eqref{eq:govern3_corr} is an equation for $\rho=\rho_0$ and not for $\rho_{\varepsilon}$, and provides a first order in $\varepsilon$ correction to the PDE~\eqref{eq:govern3}. Details of the computations of this section are reported in Appendix~\ref{sec:SM.fbot} in the Supplementary Materials.

\begin{remark}
In the same regime, a more general scaling could be considered by choosing
$
\lambda = \frac{\tilde{\lambda}}{\varepsilon}, \quad \mu = \frac{\tilde{\mu}}{\varepsilon}, \quad   \tilde{\mu}, \tilde{\lambda}=\mathcal{O}(1)
$
with $\tilde{\mu} \neq \tilde{\lambda}$, instead of~\eqref{ass:epsscal3}. The formal derivation carried out in this section could be adapted to this more general case by noting that, under this scaling, 
$
\mathcal{T}(\bnu,y):= \frac{K(y)}{\tilde{\mu}+\tilde{\lambda}}\left[\tilde{\mu}T_K(\bnu)+\tilde{\lambda}T(\bnu;y) \, \right].
$
\end{remark}

\section{Numerical simulations}
\label{sec:numsim}
In this section, we present the results of numerical simulations of the microscopic model introduced in section~\ref{sec:micro}, which is defined by the system~\eqref{eq:micro_stoch}-\eqref{eq:micro_vel}, under the parameter scaling~\eqref{ass:epsscal3}, and numerical solutions of the corresponding macroscopic model formally derived in section~\ref{sec:macro} from the phenotype-structured kinetic equation of the mesoscopic model, that is, the PDE~\eqref{eq:govern3_corr_0} -- or~\eqref{eq:govern3_corr} in the corresponding simplified scenarios previously discussed -- complemented with definitions~\eqref{def:UTKeps0red},~\eqref{def:DT}, and~\eqref{def:CT}. 

\subsection{Set-up of numerical simulations of the macroscopic model}
\label{sec:numsetupmacro}
\subsubsection*{Preliminaries} We pose the PDE~\eqref{eq:govern3_corr_0} on a bounded spatial domain $\Omega$ and complement it with the following zero-flux boundary conditions  
\begin{equation}\label{eq:cb_macro}
\begin{split}
    \Big\lbrace\rho &\Ub_{\mathcal{T}} -\varepsilon\left[\mathbb{D}_{\mathcal{T}}\nabla_\x \rho +\rho \left(\nabla_\x \cdot\mathbb{D}_{\mathcal{T}}+ \Ub_{\mathcal{T}} \nabla_\x \cdot \Ub_{\mathcal{T}} \right) \right]\\
    &-\frac{\varepsilon}{2}\left[ \left(\mathbb{C}_T-\Ub_{\mathcal{T}}\otimes\Ub_\mathcal{T}\right)\nabla_\x \rho + \left(\boldsymbol{c}_T-\Ub_{\mathcal{T}}\nabla_\x \cdot\Ub_{\mathcal{T}}\right)\rho \right]\Big\rbrace\cdot \boldsymbol{{\rm n}}=0 \qquad \text{on } \; \partial \Omega \, ,
\end{split}
\end{equation}
where $\boldsymbol{{\rm n}}$ is the outer unit normal on $\partial \Omega$. These boundary conditions can be derived from appropriate zero-flux kinetic boundary conditions as similarly done in~\citep{loy2024HJ}. We carry out numerical simulations both in a one-dimensional setting and in a two-dimensional setting. In the one-dimensional setting we take ${\bf x} \equiv x \in \Omega := [-L_m,L_M]$, while in the two-dimensional setting we take ${\bf x} \equiv (x_1,x_2)^\intercal \in \Omega :=  [0,L_M]\times[-L_m,L_M]$, where $L_m, L_M \in \mathbb{R}^+$ with $L_m<L_M$. In particular, coherently with the experimental set-up employed in~\cite[Figure 1A]{goodman1989e8}, we take $L_m=0.06$cm and $L_M=0.06$cm. As for the scaling parameter $\varepsilon$ in~\eqref{eq:govern3_corr}, in the one-dimensional setting we explore the effect of considering different values of this parameter by taking $\varepsilon \in \{10^{-2}, 10^{-3}, 10^{-4} \}$, while in the two-dimensional setting simulations are carried out with $\varepsilon=10^{-3}$. 

Moreover, consistently with subsection~\ref{mesomacromodels}, we focus on the case where the phenotypic transition kernel $K[\mathcal{S}^\ddagger]({y}|y')$ satisfies assumption~\eqref{ass:Kclosure} -- i.e. $K[\mathcal{S}^\ddagger]({y}|y') \equiv K[\mathcal{S}^\ddagger](y)$ -- so that definitions~\eqref{def:UTKeps0red},~\eqref{def:DT}, and~\eqref{def:CT} hold.

Finally, to reproduce the experimental set-up of Figure 1A in~\citep{goodman1989e8}, an illustration of which is given in Figure~\ref{fig:ex} in the Supplementary Materials, assuming the concentrations of the adhesive glycoproteins laminin and fibronectin and the ECM density in the system to be fixed, we introduce the non-negative, real functions $C_L({\bf x})$ and $C_F({\bf x})$ to model the concentrations  (in non-dimensional form) of laminin and fibronectin, respectively, and the non-negative, real function $M({\bf x})$ to model the ECM density (also in non-dimensional form). We then use these functions to identify $\mathcal{S}$, $\mathcal{S}^\dagger$, and $\mathcal{S}^\ddagger$ when modelling cell reorientation and phenotypic changes, as detailed below.

\subsubsection*{Modelling cell reorientation} We assume, for simplicity, that the reorientation of cell polarity is solely dependent on the sensed ECM density; hence, we set $\mathcal{S}(t,{\bf x}) \equiv M({\bf x})$ in \eqref{def:T_fact} and \eqref{def:B}. Note that, coherently with assumptions~\eqref{ass:constenvfact}, this implies that $\mathcal{S}$ is constant in time. Moreover,  we assume that the direction of polarity of the cells is biased towards the direction where they sense a higher ECM density, phenomenon known as haptotaxis~\citep{smith2004measurement}.  
Under these assumptions, we define the function $b$ in~\eqref{def:B} as
\begin{equation}\label{def:TrM:sim}
b[\mathcal{S}](r,\hv') \equiv b[M](r,\hv') := M({\bf x}+r{\hv'}). 
\end{equation}

Moreover, we let the cell post-reorientation speed be determined by the amount and strength of adhesion sites sensed by the cells, which we assume to be directly linked to the concentrations of laminin and fibronectin. Hence, we set $\mathcal{S}^\dagger(t,{\bf x}) \equiv \left(C_L({\bf x}),C_F({\bf x})\right)$ in~\eqref{def:T_fact} and~\eqref{def:Psi}, which implies, coherently with assumptions~\eqref{ass:constenvfact}, that $\mathcal{S}^\dagger$ is constant in time. In particular,  we assume the probability distribution $\psi[\mathcal{S}^\dagger](v) \equiv \psi[C_L,C_F](v)$ in~\eqref{def:Psi} to be such that the mean post-reorientation speed acquired by cells at position ${\bf x}$, which is defined via~\eqref{ass:normpsi}, is 
\begin{equation}\label{def:vphi:sim}
u_\psi[\mathcal{S}^\dagger] \equiv u_\psi[C_L,C_F] = w_L[C_L,C_F]\, v^L + w_F[C_L,C_F] \,v^F\,.
\end{equation}
Here $v^F,v^L \in \mathbb{R}^+_0$ with $0\leq v^F \leq v^L \leq V_\text{max}$ denote, respectively, the mean post-reorientation speeds preferentially acquired by cells adhering to laminin and fibronectin. The assumption that $v^F \leq v^L$ follows from experimental results in~\citep{goodman1989e8}. Specifically, in order to match the corresponding cell velocities measured in~\citep{goodman1989e8},
we take
\begin{equation}\label{eq:vLvFval}
v^L=2\text{cm/h} \quad \text{and} \quad v^F=0.6\text{cm/h} \, .
\end{equation}
The weights $w_L[C_L,C_F]$ and $w_F[C_L,C_F]$ in~\eqref{def:vphi:sim} provide a measure of the adhesive strength of cells to laminin and fibronectin, and we define them as the local fractions of these two proteins, i.e.
\begin{equation}\label{def:w:sim}
w_L[C_L,C_F] := \frac{C_L}{C_L+C_F} \, , \qquad w_F[C_L,C_F] := \frac{C_F}{C_L+C_F}\,.
\end{equation}

Furthermore, we define the probability distribution $\psi[\mathcal{S}^\dagger](v) \equiv \psi[C_L,C_F](v)$ as a Dirac delta distribution centered in $v=u_\psi[C_L,C_F]$, i.e.
$$
\psi[C_L,C_F](v) := \delta \left(\, v- u_\psi[C_L,C_F]\,\right)\,.
$$

Finally, in line with assumptions~\eqref{ass:Ry}, we define the sensing radius of cells in the phenotypic state $y$ as
\begin{equation}\label{def:R}
R(y) := R_{\rm min} + y\left( \, R_{\rm max}-R_{\rm min}\,\right) \, .
\end{equation}
Consistently with~\citep{sen2009matrix}, where the authors observed cells with radii as small as one twentieth and as large as one half of the width of each stripe, we set {$R_\text{min}=5\times10^{-4}$cm} and $R_\text{max}=5\times10^{-3}$cm (see also Figure~\ref{fig:ex} in the Supplementary Materials for a visual comparison with the physical domain).

\subsubsection*{Modelling phenotypic changes} We assume phenotypic changes undergone by the cells to be influenced by the local environmental conditions, which are determined by the concentrations of laminin and fibronectin. Hence, we set
$\mathcal{S}^\ddagger(t,{\bf x})\equiv \left(C_L({\bf x}),C_F({\bf x})\right)$ in the phenotypic transition kernel, that is, $K[\mathcal{S}^\ddagger](y) \equiv K[C_L,C_F](y)$. Again, note that this implies that $\mathcal{S}^\ddagger$ is constant in time, which is coherent with assumptions~\eqref{ass:constenvfact}.

We let $K(y)$ be such that the phenotypic state in which cells enter, on average, at position ${\bf x}$ as a result of phenotypic changes, which is defined via~\eqref{ass:K1}, is
\begin{equation}\label{def:yK:sim}
y_K[\mathcal{S}^\ddagger] \equiv y_K[C_L,C_F] = w_L[C_L,C_F]\, y^L_K + w_F[C_L,C_F] \,y^F_K\,.
\end{equation}
Here $y^L_K, y^F_K \in \mathbb{R}^+_0$ with $0\leq y^F_K \leq y^L_K\leq 1$ are, respectively, the phenotypic states in which, on average, cells are preferentially led to enter by phenotypic changes driven by signalling cascades triggered by laminin and fibronectin binding to cell surface receptors, and the assumption that $y^F_K \leq y^L_K$ follows from experimental evidence on cytoskeletal organisation reported in~\citep{goodman1989e8}. Moreover, the weights $w_L[C_L,C_F]$ and $w_F[C_L,C_F]$, which are defined via~\eqref{def:w:sim}, provide a measure of the strength of the mechanotransductive signals of laminin and fibronectin.

Specifically, in~\eqref{def:yK:sim} we choose
\begin{equation}\label{eq:caseAB}
y^F_K = 0 \, 
\end{equation}
to model the scenario where cells on fibronectin are induced to enter phenotypic states corresponding to a stiffer cytoskeleton, and thus a smaller sensing radius (cf. definition~\eqref{def:R}), which is consistent with what reported in~\citep{goodman1989e8}. 

Furthermore, we investigate two cases for the phenotypic state in which, on average, cells are preferentially led to enter by phenotypic changes driven by signalling cascade triggered by laminin in~\eqref{def:yK:sim}:
 \begin{itemize}
\item the case where
\begin{equation}\label{eq:caseA}
 y^L_K=0 \, ,
\end{equation}
which models the scenario where cells on laminin are induced to enter phenotypic states corresponding to a stiffer cytoskeleton, and thus a smaller sensing radius (cf. definition~\eqref{def:R});
\item the case where
\begin{equation}\label{eq:caseB}
y^L_K=1 \, ,
\end{equation}
which models the scenario, consistent with experimental evidence of~\citep{goodman1989e8}, where cells on laminin are induced to enter phenotypic states corresponding to a more flexible cytoskeleton, and thus a larger sensing radius (cf. definition~\eqref{def:R}).
\end{itemize}

Moreover, building on the modelling strategies presented in~\citep{conte2023non,loy2020modelling}, we consider two possible definitions of the phenotypic transition kernel $K[C_L,C_F](y)$:
\begin{itemize}
\item a Dirac delta distribution centered in $y_K[C_L,C_F]$, i.e.
\begin{equation}\label{def:K:dirac}
K[C_L,C_F](y) := \delta \left(\, y-y_{K}[C_L,C_F]\,\right)\,;
\end{equation}
\item a unimodal von Mises distribution with concentration parameter $k_y$ (specifically, we take $k_y=2$) and circular mean $y_K[C_L,C_F]$, i.e.
\begin{equation}\label{def:K:VM}
K[C_L,C_F](y) := 
\frac{1}{I_0(k_y)} \exp \Big[ \, k_y \cos\left(\, 2\pi (y-y_K[C_L,C_F])\,\right)\Big] \, ,
\end{equation}
where $I_0$ is the modified Bessel function of the first kind.
\end{itemize}
Definition~\eqref{def:K:dirac} translates into mathematical terms the idea that cells exposed to the same environmental conditions are led by phenotypic changes to enter exactly the same phenotypic state (i.e. there is zero phenotypic variance). On the other hand, definition~\eqref{def:K:VM} takes into account the fact that, even if the environmental conditions are the same, due to variability in intra-cellular regulatory dynamics amongst cells, there can be variability in the phenotypic state acquired by the cells undergoing phenotypic changes (i.e. there is non-zero phenotypic variance). 

For the sake of brevity, in the remainder of this section:
\begin{itemize}
\item in the light of definitions~\eqref{def:R} and~\eqref{eq:caseAB}, we will refer to the scenario corresponding to definition~\eqref{eq:caseA} as the case of ``small sensing radius on fibronectin and laminin'', while the scenario corresponding to definition~\eqref{eq:caseB} will be referred to as the case of ``small sensing radius on fibronectin and large sensing radius on laminin'';
\item the scenario corresponding to definition~\eqref{def:K:dirac} will be referred to as the case of ``zero phenotypic variance'', while we will refer to the scenario corresponding to definition~\eqref{def:K:VM} as the case of ``non-zero phenotypic variance''.
\end{itemize}

Finally, we remark that in the case of definition~\eqref{def:K:dirac} we directly solve the reduced PDE~\eqref{eq:govern3_corr}, which descends from~\eqref{eq:govern3_corr_0}.

\begin{remark}
Note that, since numerical simulations are carried out in the framework of assumptions~\eqref{ass:Kclosure} and~\eqref{ass:constenvfact}, and thus the normalised distribution of the cell population in the phenotype space is  stationary and coincides with $K[\mathcal{S}^\ddagger](y)$ (cf. \eqref{eq:p0.n0}), in the case of small sensing radius on fibronectin and laminin the mean phenotypic state of the cell population is expected to be $y_K^F$ over the fibronectin stripe and $y_K^L$  over the laminin stripes. Moreover, in the zero phenotypic variance case all cells are expected to be in the phenotypic state $y_K^F$ over the fibronectin stripe and in the phenotypic state $y_K^L$ over the laminin stripes. On the other hand, in the non-zero phenotypic variance case, the phenotypic states of the cells are expected to be scattered around $y_K^F$ over the fibronectin stripe and around $y_K^L$ over the laminin stripes.
\end{remark}

\subsubsection*{Definitions of the laminin and fibronectin concentrations and the ECM density} 
Consistently with the experimental set-up employed in~\cite[Figure 1A]{goodman1989e8}, in the two-dimensional setting, we let laminin and fibronectin be distributed along parallel stripes which run along the $x_2$ direction.  
In the one-dimensional setting, we let laminin be distributed along the stripe running along the $x$ direction.
Full details on the definitions of the laminin and fibronectin concentrations and the ECM density in the one- and two-dimensional settings are provided in Section~\ref{sec:numdets}, and illustrated in Figure~\ref{fig:ex}, in the Supplementary Materials.

\subsubsection*{Initial conditions} 
We complement the PDE~\eqref{eq:govern3_corr_0} with initial conditions that are in line with the experimental set-up employed in~\cite[Figure 1A]{goodman1989e8}, see also Figure~\ref{fig:ex} in the Supplementary Materials. Specifically, in the two-dimensional setting we use the following initial condition
\begin{equation}\label{ic:rho:sim}
\rho^0(x_1, x_2) := 
\begin{cases}
\bar{\rho}^0\left(1-\left(\dfrac{-L_m+2l_m -x_2}{l_m}\right)^2\right) &-L_m +l_m\leq x_2< -L_m +2l_m \, , \\
\bar{\rho}^0 & -L_m +2l_m\leq  x_2\leq -l_m \, , \\
\bar{\rho}^0\left(1-\left(\dfrac{x_2+l_m}{l_m}\right)^2\right) &-l_m<x_2\leq 0 \, , \\
0 &x_2<-L_m +l_m \;\; \vee \;\; x_2>0 \, ,
\end{cases}
\end{equation}
for all $x_1\in[0,L_M]$, where $\bar{\rho}^0 \in \mathbb{R}^+$ and $l_m \in (0, L_m)$. Similarly, in the one-dimensional setting we use the following initial condition
\begin{equation}\label{ic:rho:sim1D}
\rho^0(x) := 
\begin{cases}
\bar{\rho}^0\left(1-\left(\dfrac{-L_m+2l_m -x}{l_m}\right)^2\right) &-L_m +l_m\leq x< -L_m +2l_m \, , \\
\bar{\rho}^0 &-L_m +2l_m\leq  x\leq -l_m \, , \\
\bar{\rho}^0\left(1-\left(\dfrac{x+l_m}{l_m}\right)^2\right) &-l_m<x\leq 0 \, , \\
0 &x<-L_m +l_m \;\;\vee\;\; x>0 \, .
\end{cases}
\end{equation}
In particular, we choose $l_m=0.01$cm and $\bar{\rho}^0=2\times10^3$cells/cm$^2$. 
\\

\paragraph{Numerical methods} Numerical solutions are constructed on a uniform grid by using a mixed finite-difference and finite-volume scheme, as detailed in Appendix~\ref{sec:num:method} in the Supplementary Materials. The numerical scheme is implemented in {\sc MATLAB}\textsuperscript{\textregistered} and the source code is available on GitHub\footnote{\url{https://github.com/ChiaraVilla/LorenziEtAl2026Modelling}\label{github}}.

\subsection{Set-up of numerical simulations of the microscopic model}
\label{sec:numsetupmicro}
For consistency with the macroscopic model~\eqref{eq:govern3_corr_0}, which is formally obtained under the parameter scaling~\eqref{ass:epsscal3} and making assumption~\eqref{ass:Kclosure}, numerical simulations of the microscopic model defined by the system~\eqref{eq:micro_stoch}-\eqref{eq:micro_vel} are carried out letting assumptions~\eqref{ass:epsscal3} and~\eqref{ass:Kclosure} hold and using a set-up corresponding to the one employed for the macroscopic model. To carry out numerical simulations, we implement a Nanbu-Babovski Monte Carlo scheme, as similarly done in~\citep{conte2023non,loy2021label}. We use a sufficiently large number of particles (i.e. $10^6$ particles) and a sufficiently small time-step (i.e. $\Delta t=10^{-5}$), so as to ensure that the implemented Monte Carlo scheme can provide a good numerical approximation of the solution to the phenotype-structured kinetic equation~\eqref{eq:Boltz_coll_2:strong}. The Monte Carlo scheme is implemented in {\sc MATLAB}\textsuperscript{\textregistered} and the source code is available on GitHub\footref{github}.

\subsection{Main results of numerical simulations}

\subsubsection{Comparison between the microscopic and the macroscopic models}
\begin{figure}[h!]
\centering
\includegraphics[width=.95\linewidth]{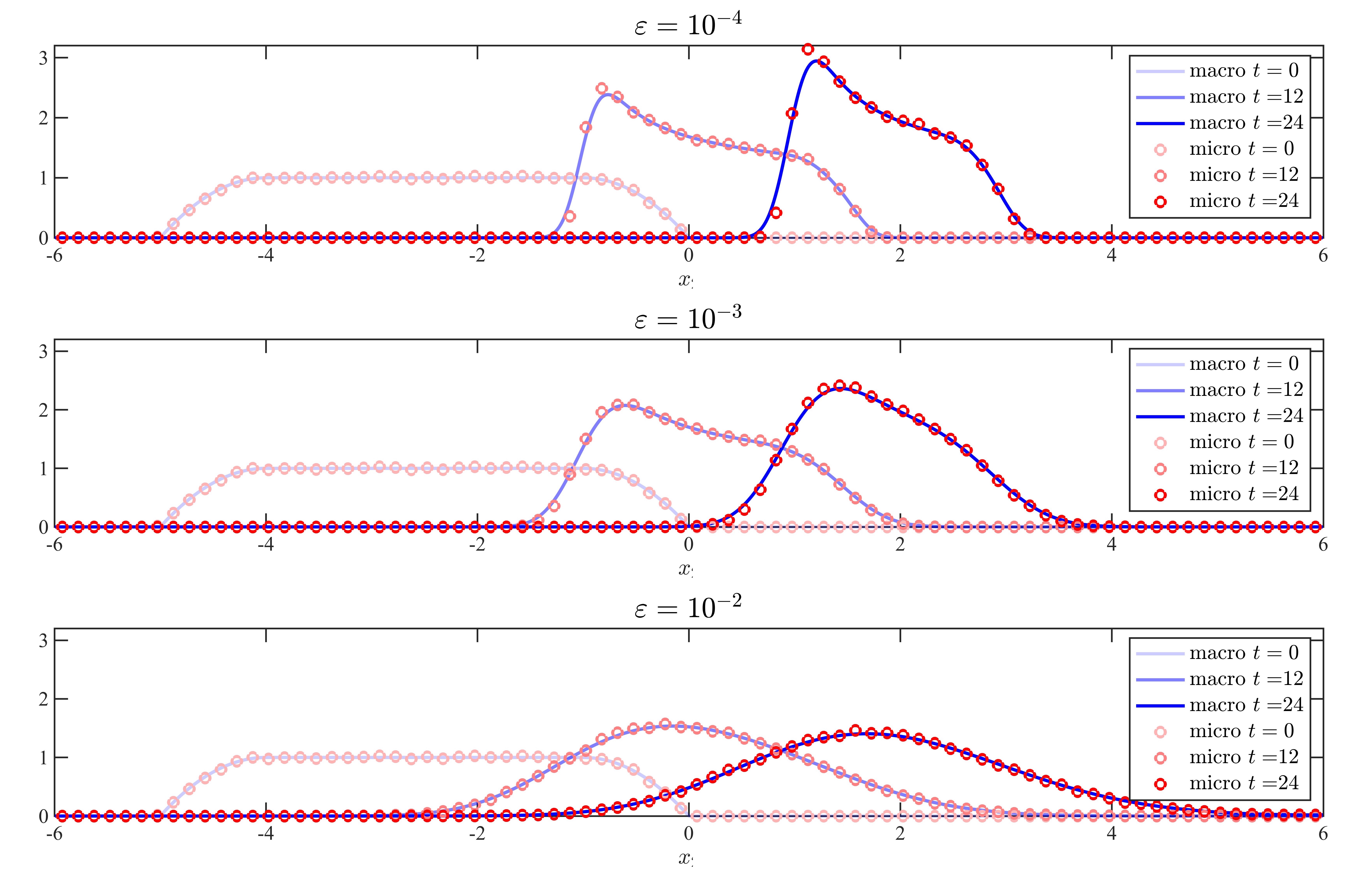}
\caption{\label{fig:compare1d} \textbf{Comparison between the microscopic and the macroscopic models in 1D.}
Plots of the cell number density, normalised with respect to $\bar{\rho}^0$, obtained from 1D numerical simulations of the microscopic model (red dots) and the numerical solution of the corresponding macroscopic model (blue lines), also normalised with respect to $\bar{\rho}^0$, at $t=0$h, $t=12$h, and $t=24$h (decreasing transparency levels). The microscopic model consists of the system \eqref{eq:micro_stoch}-\eqref{eq:micro_vel} subject to the parameter scaling~\eqref{ass:epsscal3}, while the macroscopic model comprises the PDE~\eqref{eq:govern3_corr} complemented with definitions~\eqref{def:UTKeps0red} and~\eqref{def:DT}. Numerical simulations are carried out under the set-up detailed in subsections~\ref{sec:numsetupmacro} and \ref{sec:numsetupmicro}, in the one-dimensional setting, with definition~\eqref{def:K:dirac}, assumption~\eqref{eq:caseB}, and the scaling parameter $\varepsilon=10^{-4}$ (top row), $\varepsilon=10^{-3}$ (middle row) or $\varepsilon=10^{-2}$ (bottom row). The variable $x$ in the plots is in units of $10^{-2}$cm.  
}
\end{figure}

\begin{figure}[h!]
\centering
  \includegraphics[width=\linewidth]{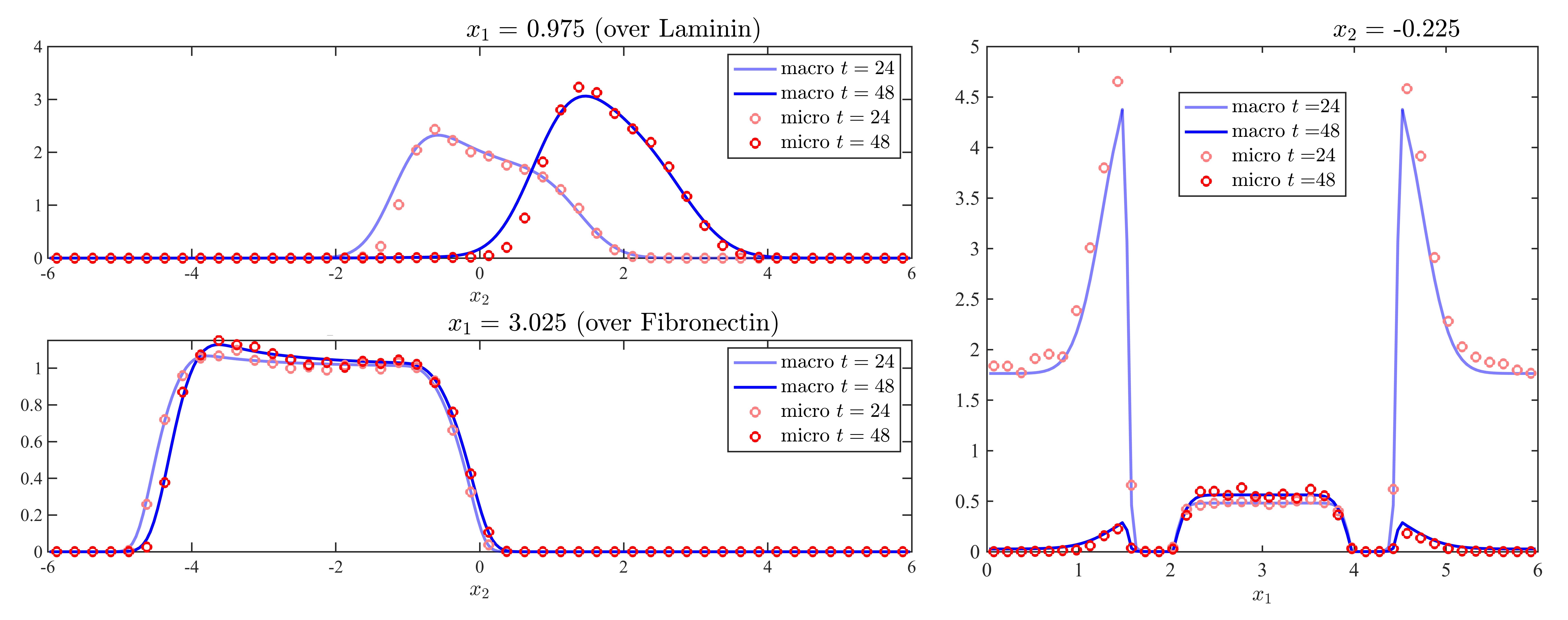}
\caption{\label{fig:compare2d} \textbf{Comparison between the microscopic and the macroscopic models in 2D.}
Plots of the cross-sections of the cell number density, normalised with respect to $\bar{\rho}^0$, obtained from 2D numerical simulations of the microscopic model (red dots) and the numerical solution of the corresponding macroscopic model (blue lines), also normalised with respect to $\bar{\rho}^0$, at $t=0$h, $t=24$h, and $t=48$h (decreasing transparency levels).  Longitudinal cross-sections at $x_1=0.975\times 10^{-2}$cm (top) and $x_1=3.025\times 10^{-2}$cm (bottom) -- i.e. over laminin and fibronectin, respectively -- are displayed on the left, while transversal cross-sections at $x_2=-0.225\times 10^{-2}$cm are displayed on the right. The full plot of the numerical solution of the macroscopic model at $t=48$h is displayed in Figure~\ref{fig:stripes-dd}B. The microscopic model consists of the system \eqref{eq:micro_stoch}-\eqref{eq:micro_vel} subject to the parameter scaling~\eqref{ass:epsscal3}, while the macroscopic model comprises the PDE~\eqref{eq:govern3_corr} complemented with definitions~\eqref{def:UTKeps0red} and~\eqref{def:DT}. Numerical simulations are carried out under the set-up detailed in subsections~\ref{sec:numsetupmacro} and \ref{sec:numsetupmicro}, in the two-dimensional setting, with definition~\eqref{def:K:dirac}, assumption~\eqref{eq:caseB}, and the scaling parameter $\varepsilon=10^{-3}$. The variables $x_1$ and $x_2$ in the plots are in units of $10^{-2}$cm.}
\end{figure}

The plots in Figure~\ref{fig:compare1d} present a comparison between the cell number density obtained from numerical simulations of the microscopic model and the numerical solution of the macroscopic model in the one-dimensional setting corresponding to the set-up described in subsections~\ref{sec:numsetupmacro} and \ref{sec:numsetupmicro}, for different values of the scaling parameter $\varepsilon$. These plots demonstrate that there is an excellent quantitative agreement between the two models, for all the values of the scaling parameter $\varepsilon$ here considered. Note that, as it can be expected from the form of the PDE~\eqref{eq:govern3_corr}, larger values of $\varepsilon$ lead to broader spatial distributions of cells.

An excellent quantitative agreement between the microscopic and the macroscopic models is also observed in the two-dimensional setting corresponding to the set-up described in subsections~\ref{sec:numsetupmacro} and \ref{sec:numsetupmicro}, as shown by the plots in Figure~\ref{fig:compare2d}, which display cross-sections of the cell number density resulting from numerical simulations of the microscopic model alongside cross-sections of the numerical solution of the corresponding macroscopic model, when $\varepsilon=10^{-3}$. Note that the agreement between the two models deteriorates slightly in the regions where the cell number density undergoes sharper changes, as it can be expected due to the histogram method through which the cell number density for the microscopic model is reconstructed from the results of Monte Carlo simulations.

Taken together, these results validate the formal procedures employed to derive first the mesoscopic model, defined by the phenotype-structured kinetic equation~\eqref{eq:Boltz_coll_2:strong}, and then the macroscopic model, at least in the case where $K$ is defined via~\eqref{def:K:dirac}, yielding the PDE~\eqref{eq:govern3_corr} complemented with definitions~\eqref{def:UTKeps0red} and~\eqref{def:DT}, from the microscopic model consisting of the system \eqref{eq:micro_stoch}-\eqref{eq:micro_vel} in the regime~\eqref{ass:epsscal3}.

\subsubsection{Comparison between the macroscopic model and experimental results}\label{sec:num:res:2D}
\begin{figure}[htb]
\centering
\includegraphics[width=\linewidth]{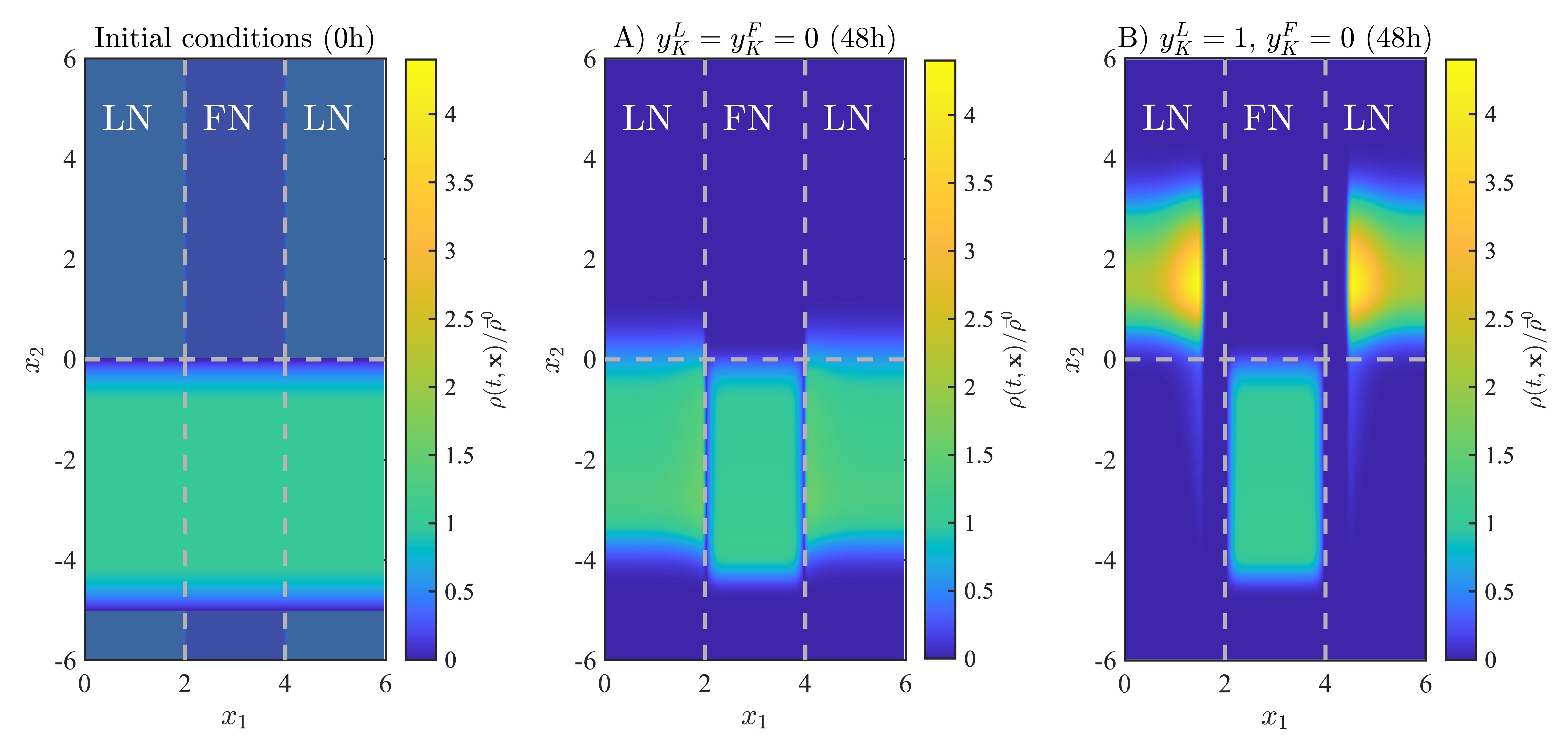}
\caption{\label{fig:stripes-dd} \textbf{Numerical solutions of the macroscopic model in 2D under zero phenotypic variance.} 
Plots of the numerical solution of the macroscopic model, normalised with respect to $\bar{\rho}^0$, at $t=0$h (left column) and at $t=48$h (central and right columns). The macroscopic model comprises the PDE~\eqref{eq:govern3_corr} complemented with definitions~\eqref{def:UTKeps0red} and~\eqref{def:DT}. Numerical simulations are carried out under the set-up detailed in subsection~\ref{sec:numsetupmacro}, in the two-dimensional setting, with definition~\eqref{def:K:dirac},  assumption~\eqref{eq:caseA} (central column) or assumption~\eqref{eq:caseB} (right column), and the scaling parameter $\varepsilon=10^{-3}$. The variables $x_1$ and $x_2$ in the plots are in units of $10^{-2}$cm. Laminin and fibronectin stripes are highlighted by ${\rm LN}$ and ${\rm FN}$, respectively.
}
\end{figure}

\begin{figure}[htb!]
\centering
\includegraphics[width=\linewidth]{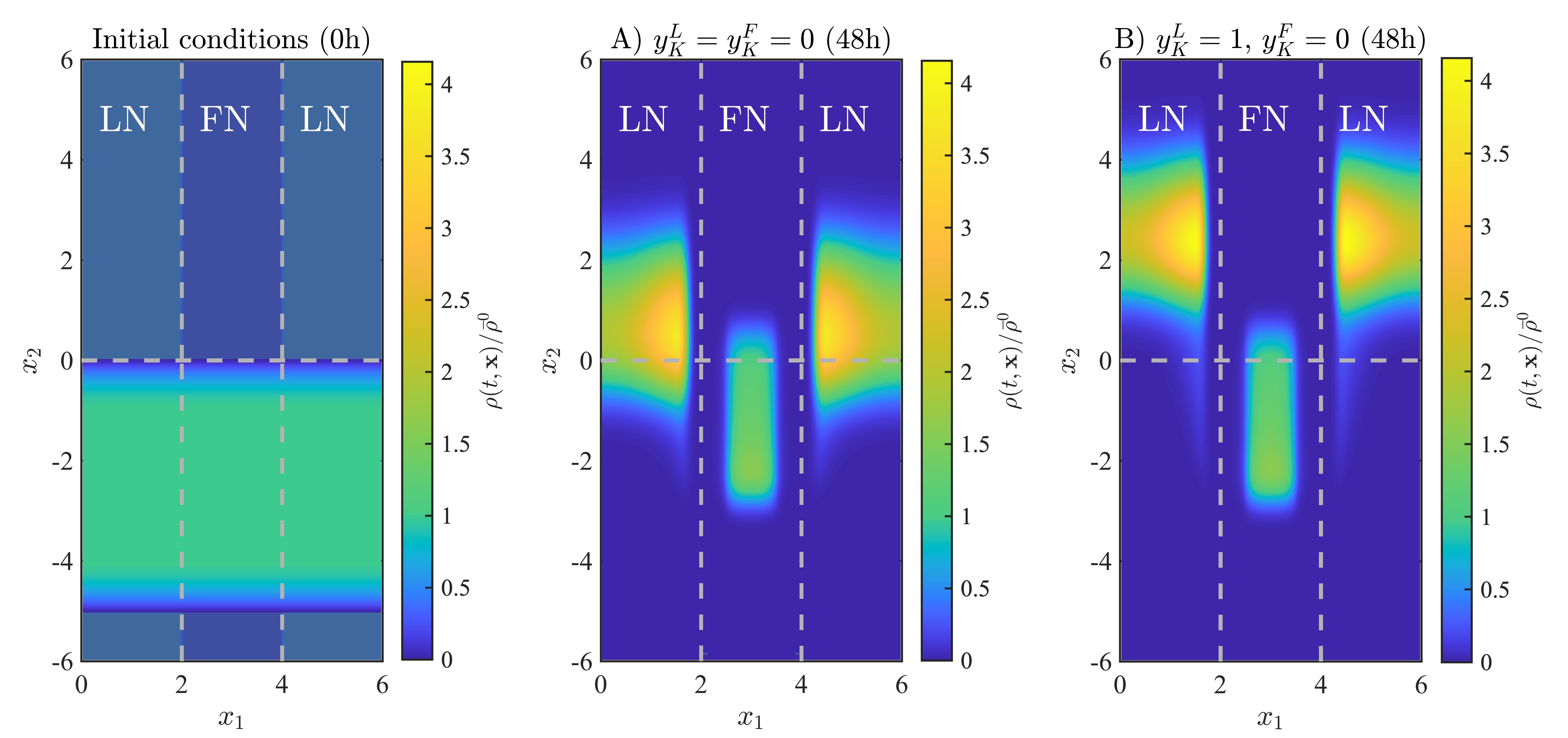}
\caption{\label{fig:stripes-vm} \textbf{Numerical solutions of the macroscopic model in 2D under non-zero phenotypic variance.} 
Plots of the numerical solution of the macroscopic model, normalised with respect to $\bar{\rho}^0$, at $t=0$h (left column) and at $t=48$h (central and right columns). The macroscopic model comprises the PDE~\eqref{eq:govern3_corr_0} complemented with definitions~\eqref{def:UTKeps0red},~\eqref{def:DT}, and~\eqref{def:CT}. Numerical simulations are carried out under the set-up detailed in subsection~\ref{sec:numsetupmacro}, in the two-dimensional setting, with definition~\eqref{def:K:VM},  assumption~\eqref{eq:caseA} (central column) or assumption~\eqref{eq:caseB} (right column), and the scaling parameter $\varepsilon=10^{-3}$. The variables $x_1$ and $x_2$ in the plots are in units of $10^{-2}$cm. Laminin and fibronectin stripes are highlighted by ${\rm LN}$ and ${\rm FN}$, respectively.
}
\end{figure}

The plots in Figures~\ref{fig:stripes-dd} and~\ref{fig:stripes-vm} display numerical solutions of the macroscopic model defined by the PDE~\eqref{eq:govern3_corr_0} complemented with definitions~\eqref{def:UTKeps0red},~\eqref{def:DT}, and~\eqref{def:CT}, under the set-up detailed in subsection~\ref{sec:numsetupmacro}, in the two-dimensional setting mimicking the experimental set-up of the stripe migration assay of~\citep{goodman1989e8}. In more detail, Figure~\ref{fig:stripes-dd} refers to the case of zero phenotypic variance (i.e. when the phenotypic transition kernel, $K$, is defined via~\eqref{def:K:dirac}), while Figure~\ref{fig:stripes-vm} refers to the case of non-zero phenotypic variance (i.e. when $K$ is defined via~\eqref{def:K:VM}). Moreover, the results in Figures~\ref{fig:stripes-dd}A and~\ref{fig:stripes-vm}A are for the case of small sensing radius on fibronectin and laminin (i.e. when the phenotypic states in which, on average, cells are preferentially led to enter by signalling cascades triggered by fibronectin, $y^F_K$, and laminin, $y^L_K$, are defined via~\eqref{eq:caseAB} and~\eqref{eq:caseA}). On the other hand, the results in Figures~\ref{fig:stripes-dd}B and~\ref{fig:stripes-vm}B are for the case of small sensing radius on fibronectin and large sensing radius on laminin (i.e. when $y^F_K$ is still defined via~\eqref{eq:caseAB} but $y^L_K$ is defined via~\eqref{eq:caseB} instead).

The numerical solutions displayed in Figure~\ref{fig:stripes-dd} and Figure~\ref{fig:stripes-vm} show that there is accumulation of cells on the laminin stripes. This is due to the fact that, since cell reorientation is driven by non-local sensing of the surrounding environment, cells near the boundary between the fibronectin stripe and the laminin stripes will eventually move to the laminin stripes, where their motion can occur at a higher speed (cf. assumption~\eqref{eq:vLvFval}). However, in the case of small sensing radius on fibronectin and large sensing radius on laminin, which is the one consistent with the experimental study in~\citep{goodman1989e8} (cf. Figure~\ref{fig:stripes-dd}B and Figure~\ref{fig:stripes-vm}B), accumulation of cells over laminin stripes is more pronounced than in the case of small sensing radius on fibronectin and laminin (cf. Figure~\ref{fig:stripes-dd}A and Figure~\ref{fig:stripes-vm}A). Furthermore, the part of the cell population adhering to laminin undergoes appreciably faster migration than the part of the cell population adhering to fibronectin, an emergent property that was also experimentally observed in~\citep{goodman1989e8}.

The numerical solutions displayed in Figure~\ref{fig:stripes-dd} and Figure~\ref{fig:stripes-vm} also indicate that migration of the cell population is slightly faster, both on fibronectin and on laminin, when the phenotypic variance in non-zero (cf. Figure~\ref{fig:stripes-vm}) than when there is zero phenotypic variance (cf. Figure~\ref{fig:stripes-dd}), as it is especially evident in the case of small sensing radius on fibronectin and laminin (cf. Figure~\ref{fig:stripes-dd}A and Figure~\ref{fig:stripes-vm}A). This can be explained by noting that in this case cells will have a tendency to enter the phenotypic state $y=0$ (which corresponds to a stiff cytoskeleton and the smallest sensing radius, $R_{\rm min}$) both on laminin and on fibronectin. Hence, phenotypic variance will be the sole catalyst for the appearance of cells in phenotypic states $y>0$ (which correspond to a more flexible cytoskeleton and a larger sensing radius). Cells on the fibronectin stripe which are in such phenotypic states will then reach more easily the laminin stripes, where they will move, on average, at a higher speed, thus promoting the emergence of faster migration of the cell population. This is slightly less apparent in the case of small sensing radius on fibronectin and large sensing radius on laminin. In fact, in this case, if the phenotypic variance is non-zero then the sensing radii of the cells will be more uniformly distributed between $R_{min}$ and $R_{max}$ than when there is zero phenotypic variance.

\subsubsection{Comparison between the original macroscopic model and a reduced model}\label{sec:num:res:2D:SIM}
\begin{figure}[htb!]
\centering
\includegraphics[width=\linewidth]{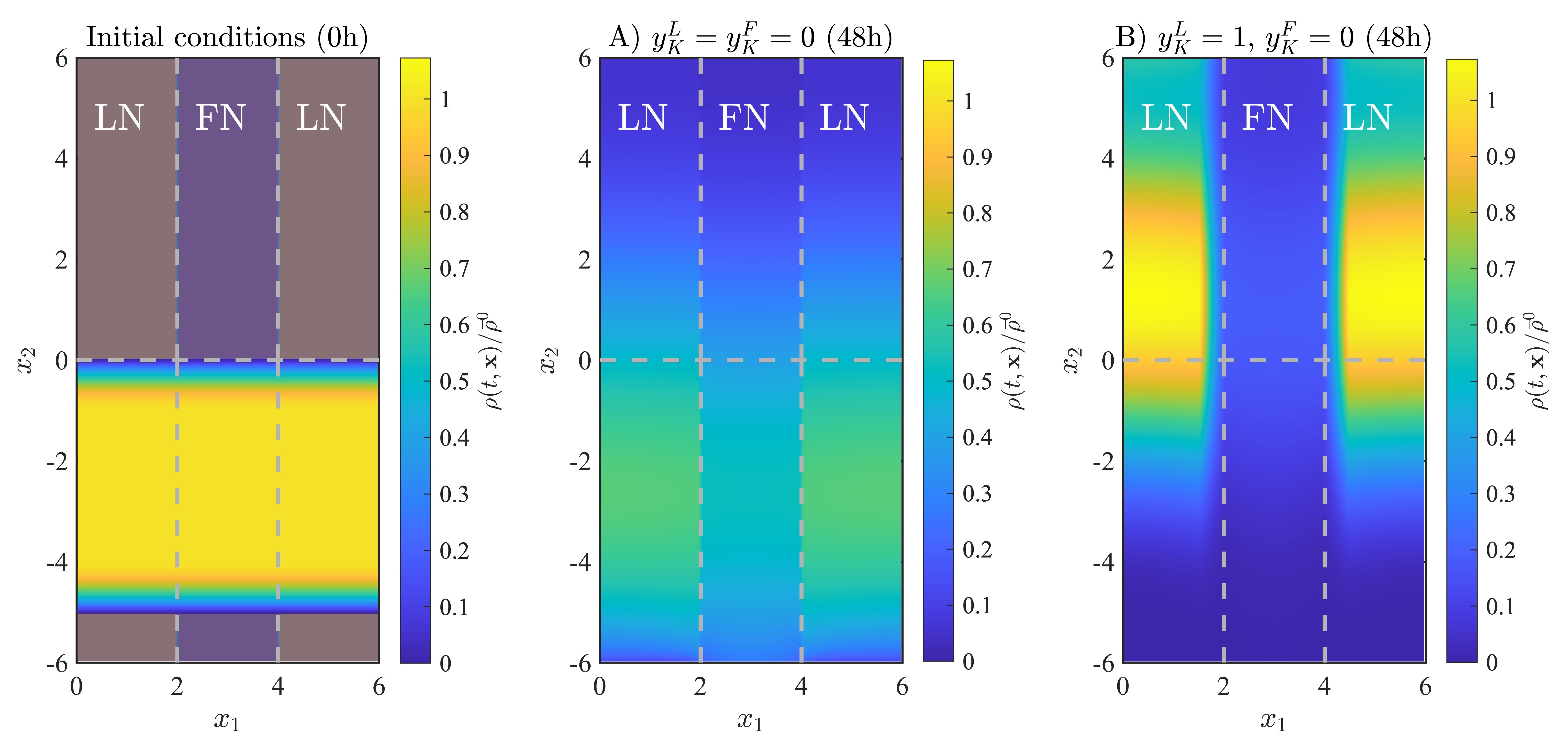}
\caption{\label{fig:stripes-dd-SIM} \textbf{Numerical solutions of the reduced macroscopic model in 2D under zero phenotypic variance.} 
Plots of the numerical solution of the reduced macroscopic model, normalised with respect to $\bar{\rho}^0$, at $t=0$h (left column) and at $t=48$h (central and right columns). The reduced macroscopic model comprises the PDE~\eqref{eq:govern3_corr_SIM} complemented with definition~\eqref{def:UTKeps0red} and subject to the boundary conditions~\eqref{eq:cb_macro_SIM}. Numerical simulations are carried out under the set-up detailed in subsection~\ref{sec:numsetupmacro}, in the two-dimensional setting, with definition~\eqref{def:K:dirac}, assumption~\eqref{eq:caseA} (central column) or assumption~\eqref{eq:caseB} (right column), and the scaling parameter $\varepsilon=10^{-1}$. The variables $x_1$ and $x_2$ in the plots are in units of $10^{-2}$cm. Laminin and fibronectin stripes are highlighted by ${\rm LN}$ and ${\rm FN}$, respectively.
}
\end{figure}

\begin{figure}[htb!]
\centering
\includegraphics[width=\linewidth]{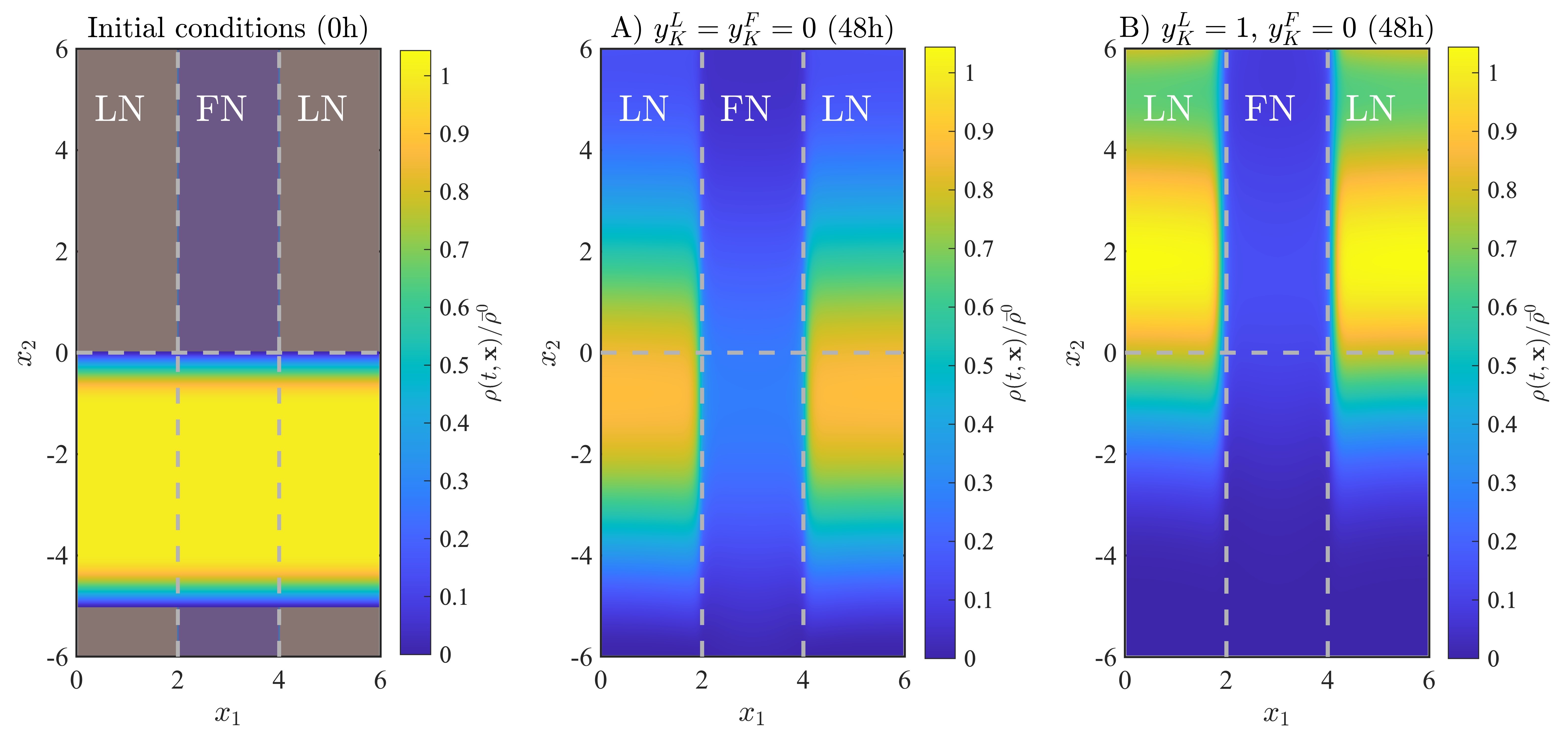}
\caption{\label{fig:stripes-vm-SIM} \textbf{Numerical solutions of the reduced macroscopic model in 2D under non-zero phenotypic variance.} 
Plots of the numerical solution of the reduced macroscopic model, normalised with respect to $\bar{\rho}^0$, at $t=0$h (left column) and at $t=48$h (central and right columns). The reduced macroscopic model comprises the PDE~\eqref{eq:govern3_corr_SIM} complemented with definition~\eqref{def:UTKeps0red} and subject to the boundary conditions~\eqref{eq:cb_macro_SIM}. Numerical simulations are carried out under the set-up detailed in subsection~\ref{sec:numsetupmacro}, in the two-dimensional setting, with definition~\eqref{def:K:dirac}, assumption~\eqref{eq:caseA} (central column) or assumption~\eqref{eq:caseB} (right column), and the scaling parameter $\varepsilon=10^{-1}$. The variables $x_1$ and $x_2$ in the plots are in units of $10^{-2}$cm. Laminin and fibronectin stripes are highlighted by ${\rm LN}$ and ${\rm FN}$, respectively.
}
\end{figure}

In the framework of the macroscopic model defined by the PDE~\eqref{eq:govern3_corr_0}, cell dynamics result from the superposition of inhomogeneous, anisotropic diffusion, with diffusion tensor $\mathbb{D}_\mathcal{T}^\varepsilon = \varepsilon\, \mathbb{D}_\mathcal{T} + \frac{\varepsilon}{2} \left(\mathbb{C}_T-\Ub_{\mathcal{T}}\otimes\Ub_\mathcal{T}\right)$, and advective transport, with velocity field$\Ub_\mathcal{T}^\varepsilon = \Ub_\mathcal{T}\left( 1 - \varepsilon \nabla_\x \cdot \Ub_\mathcal{T}\right)  - \varepsilon \nabla_\x \cdot {\mathbb{D}}_\mathcal{T} - \frac{\varepsilon}{2}\left( \boldsymbol{c}_T-\Ub_{\mathcal{T}}\nabla_\x \cdot\Ub_{\mathcal{T}}\right)$. 
In particular, the transport term represents the effect, at the cell-population level, of the interplay between directional migration, which is led by cell reorientation driven by non-local sensing of the surrounding environment, and environment-induced changes in the cytoskeletal structure. 
In order to assess the robustness of the patterns of cell migration produced by the macroscopic model, we investigate whether features qualitatively similar to those displayed by the numerical solutions presented in Figures~\ref{fig:stripes-dd} and~\ref{fig:stripes-vm} can emerge when the anisotropic diffusion term is replaced by a linear diffusion term with diffusion coefficient $\varepsilon$, and the terms of order $\varepsilon/2$ are neglected. The latter terms, which vanish in the case of $K$ defined via~\eqref{def:K:dirac}, are considerably small also in the case of $K$ defined via~\eqref{def:K:VM} -- at least for the value of the von Mises concentration parameter $k_y$ selected for the simulations -- yielding an effective contribution of order much smaller than $\varepsilon$.
Hence, we consider the following reduced macroscopic model
\begin{equation}\label{eq:govern3_corr_SIM}
\partial_t \rho + \nabla_\x \cdot \left[\rho \, \Ub_{\mathcal{T}} \left(1 - \varepsilon \, \nabla_\x \cdot \Ub_{\mathcal{T}} \right)\right]= \varepsilon \, \Delta_\x  \rho \, ,
\end{equation}
where $\Ub_{\mathcal{T}}$ is still defined via~\eqref{def:UTKeps0red}. This simplified model is obtained from the original model by replacing the variance-covariance matrix of $T$, $\mathbb{D}_{\mathcal{T}}$, in the PDE~\eqref{eq:govern3_corr_0} by the identity matrix, and neglecting terms of order $\varepsilon/2$. In analogy with what done to solve numerically the PDE~\eqref{eq:govern3_corr_0}, we pose the PDE~\eqref{eq:govern3_corr_SIM} on the spatial domain $\Omega$ and complement it with the following zero-flux boundary conditions  
\begin{equation}\label{eq:cb_macro_SIM}
    \left[\rho \Ub_{\mathcal{T}} -\varepsilon\left(\nabla_\x \rho +\rho \Ub_{\mathcal{T}} \nabla_\x \cdot \Ub_{\mathcal{T}} \right)\right]\cdot \boldsymbol{{\rm n}}=0 \qquad \text{on } \; \partial \Omega \, .
\end{equation}
Moreover, we carry out numerical simulations under the set-up detailed in subsection~\ref{sec:numsetupmacro}, in the two-dimensional setting mimicking the experimental set-up of the stripe migration assay from~\citep{goodman1989e8}. Numerical solutions are constructed using methods similar to those employed to solve numerically the PDE~\eqref{eq:govern3_corr_0}, which are detailed in Appendix~\ref{sec:num:method} in the Supplementary Materials.

Numerical solutions of the reduced macroscopic model \eqref{eq:govern3_corr_SIM} in the cases of zero and non-zero phenotypic variance are displayed in Figure~\ref{fig:stripes-dd-SIM} and Figure~\ref{fig:stripes-vm-SIM}, respectively. These numerical solutions display the same key qualitative features as those of the numerical solutions in Figures~\ref{fig:stripes-dd} and~\ref{fig:stripes-vm} (i.e. emergence of higher cell density and faster cell migration on laminin stripes), which testifies to the robustness of the patterns of cell migration produced by the macroscopic model.

\section{Discussion and research perspectives}
\label{sec:discconcs}
We developed a modelling framework for migration of heterogeneous cell populations driven by non-local environmental sensing. In the vein of previous work on phenotype-structured models of cell movement reviewed in~\citep{lorenzi2024phenotype}, this framework generalises the modelling approach proposed in~\citep{loy2020kinetic,loy2020modelling} by incorporating a continuous phenotype structure that makes it possible to take into account: inter-cellular variability in structural properties of the cytoskeleton, and thus in the length of the surface projections through which the cells sense the surrounding environment; the occurrence of environment-induced changes in structural properties of the cytoskeleton, which result in dynamical changes in the length of the cells' surface projections.  

We started by formulating a microscopic model in which single cell dynamics are described by means of stochastic processes, which represent cell movement and environment-induced changes in structural properties of the cytoskeleton. Through a limiting procedure, we formally derived a phenotype-structured kinetic equation that governs the dynamics of the cell distribution in the phase and phenotype spaces, which constitutes the mesoscopic counterpart of the microscopic model. From the mesoscopic model we formally derived a closed PDE for the cell density, which defines the corresponding macroscopic model. Such a PDE comprises an inhomogeneous, anisotropic diffusion term and an advective transport term, with the latter modelling the interplay between environment-induced changes in the cytoskeletal structure and directional migration led by cell reorientation, which is driven by non-local sensing of the surrounding environment.

We compared numerical solutions of the PDE defining the macroscopic model and the results of Monte Carlo simulations of the microscopic model, and showed that there is an excellent quantitative agreement between them, under a simulation set-up which reproduces the experimental set-up of the cell locomotion assays of~\citep{goodman1989e8}. We also showed that numerical solutions of the macroscopic model recapitulate qualitative features of experimental observations presented in~\citep{goodman1989e8} by demonstrating that the interplay between cell reorientation driven by non-local sensing of the surrounding environment and environment-induced changes in the cytoskeletal structure lead to faster migration of cells adhering to laminin stripes. To corroborate the robustness of the patterns of cell migration emerging in numerical solutions of the macroscopic model, we showed that qualitatively similar patterns are displayed by numerical solutions of a reduced model, wherein the advective transport term is left unchanged while the inhomogeneous, anisotropic diffusion term is replaced by a linear diffusion term.

We conclude with an outlook on possible research perspectives. Informed by the numerical solutions presented in Figures~\ref{fig:stripes-dd}-\ref{fig:stripes-vm-SIM}, it would be interesting to investigate the existence of travelling wave solutions for the PDE~\eqref{eq:govern3_corr_0} or the PDE~\eqref{eq:govern3_corr_SIM} in the form of travelling pulses. Moreover, we carried out numerical simulations under assumption~\eqref{ass:Kclosure} on the phenotypic transition kernel, which facilitates the explicit identification of $\Ub_{\mathcal{T}}$ and $\mathbb{D}_{\mathcal{T}}$ thanks to the result~\eqref{eq:g1explicit}, but it would also be relevant to explore how the patterns of cell migration presented here may change when this assumption is relaxed. Furthermore, while our theoretical study has eschewed specific mechanisms driving phenotypic changes, consistently with the fact that these are not detailed in the experimental study of~\citep{goodman1989e8}, modelling more precisely the effect of chemical or mechanical external stimuli that drive cytoskeletal changes would be another avenue for future research. Although, guided by the experiments of~\citep{goodman1989e8}, in this work we focused on the microscopic model under the parameter scaling~\eqref{ass:epsscal3} and the corresponding macroscopic model defined by the PDE~\eqref{eq:govern3_corr_0}, another track to follow would be to integrate the results of our simulation-based study with Monte Carlo simulations of the microscopic model under different parameter scalings, and complement these with numerical solutions of the corresponding macroscopic models. Finally, while our attention has been focused on the microscopic and macroscopic models, it would certainly be relevant to consider also the mesoscopic model defined by the phenotype-structured kinetic equation~\eqref{eq:Boltz_coll_2:strong}. In this regard, it would be interesting to investigate whether such a kinetic equation admits travelling wave solutions that exhibit phenotype structuring, whereby cells with different cytoskeleton properties dominate different parts of the wave~\citep{lorenzi2024phenotype}; for this, combining techniques similar to those employed in~\citep{bouin2019spreading,bouin2014travelling,bouin2012invasion,bouin2015propagation} and~\citep{freingruber2025trait,lorenzi2024derivation,lorenzi2022trade,lorenzi2022invasion} may prove useful. This would allow for further investigation into how phenotypic heterogeneity affects migration of cell populations driven by non-local environmental sensing.


\appendix
\renewcommand\thefigure{\thesection.\arabic{figure}} 

\section*{Supplementary Materials}
\setcounter{figure}{0}  

\section{Formal derivation of the mesoscopic model}\label{appendix}
Starting from system~\eqref{eq:micro_stoch} complemented with relations~\eqref{eq:micro_vel}, we formally derive the mesoscopic counterpart of the microscopic model presented in Section~\ref{sec:micro}. The derivation method relies on classical limiting procedures of kinetic theory for multi-agent systems, as detailed, in the context of velocity-jump processes for cell migration, in~\citep{conte2023non}.

Since the components of the quadruple $(\X_{t+\Delta t},V_{t+\Delta t},\hbV_{t+\Delta t}, Y_{t+\Delta t})$ are given by the system~\eqref{eq:micro_stoch}, for any observable $\phi: \mathbb{R}^d \times \mathcal{V} \times [0,1] \to \mathbb{R}$, in the asymptotic regime $\Delta t \to 0^+$, the expectation
\begin{equation}\label{def:expext1}
\ave{\phi\left(\X_{t},V_{t},\hbV_{t}, Y_{t}\right)} = \int_{\mathbb{R}^d} \int_{\mathcal{V}} \int_{0}^1 \phi(\x,\bnu,y) \, f(t,\x,\bnu,y) \, {\rm d}y \, {\rm d}\bnu \, {\rm d}\x
\end{equation}
formally satisfies, see for example~\citep{pareschi2013interacting}, the following differential equation
{\small
\begin{equation}\label{eq:ave_2} 
\begin{aligned}[b]
& \dfrac{{\rm d}}{{\rm d} t}\ave{\phi\left(\X_t,V_t,\hbV_t,Y_t\right)}+\nabla_\x \cdot \ave{V_t\hbV_t \, \phi(\X_t,V_t,\hbV_t,Y_t)}= \\
& \phantom{ \dfrac{{\rm d}}{{\rm d} t}\ave{\phi\left(\X_t,V_t,\hbV_t,Y_t\right)}+} \mu\ave{\phi\left(\X_t,V_t',\hbV_t',Y_t\right)-\phi\left(\X_t,V_t,\hbV_t,Y_t\right)}  \\
& \phantom{ \dfrac{{\rm d}}{{\rm d} t}\ave{\phi\left(\X_t,V_t,\hbV_t,Y_t\right)}+} +\lambda\ave{\phi\left(\X_t,V_t,\hbV_t,Y'_t\right)-\phi\left(\X_t,V_t,\hbV_t,Y_t\right)}
\end{aligned}
\end{equation}
}
with
{\footnotesize
\begin{eqnarray}\label{def:expext2}
\ave{\phi\left(\X_t,V_t',\hbV_t',Y_t\right)}&=&\int_{\mathbb{R}^d}\int_{\mathcal{V}}\int_0^1\int_{\mathcal{V}} \Psi[\cS^\dagger]({v'};\hv',y) \, \mathcal{B}[\cS](\hv';y) \, \phi(\x,\bnu',y) \, {\rm d}\bnu' \, f(t,\x,\bnu,y) \, {\rm d}y \, {\rm d}\bnu \, {\rm d}\x \nonumber \\
&=& \int_{\mathbb{R}^d}\int_{\mathcal{V}}\int_0^1\int_{\mathcal{V}} T[\mathcal{S},\mathcal{S}^\dagger](\bnu';y) \, \phi(\x,\bnu',y) \, {\rm d}\bnu' \, f(t,\x,\bnu,y) \, {\rm d}y \, {\rm d}\bnu \, {\rm d}\x \, ,
\end{eqnarray}
}
where the kernel $T$ is defined via~\eqref{def:T_fact}, and
{\small
\begin{equation}\label{def:expext3}
\ave{\phi\left(\X_t,V_t,\hbV_t,Y'_t\right)}=\int_{\mathbb{R}^d}\int_{\mathcal{V}}\int_0^1\int_{0}^1 K[\mathcal{S}^\ddagger](y'|y) \, \phi(\x,\bnu,y') \, {\rm d}y' \, f(t,\x,\bnu,y) \, {\rm d}y \, {\rm d}\bnu \, {\rm d}\x \, ,
\end{equation}
}
with the kernel $K$ satisfying assumptions~\eqref{ass:K1}. By using \eqref{def:expext1}, \eqref{def:expext2}, and \eqref{def:expext3}, we rewrite \eqref{eq:ave_2} as
{\footnotesize
$$
\begin{aligned}[b]
&\dfrac{{\rm d}}{{\rm d}t}\int_{\mathbb{R}^d} \int_{\mathcal{V}}\int_0^1\phi(\x,\bnu,y) \, f(t,\x,\bnu,y) \, {\rm d}y \, {\rm d}\bnu \, {\rm d}\x +\int_{\mathbb{R}^d} \nabla_\x \cdot \int_{\mathcal{V}}\int_0^1 \phi(\x,\bnu,y) \, v \hv \, f(t,\x,\bnu,y) \, {\rm d}y \, {\rm d}\bnu \,  {\rm d}\x =\\
&\qquad\quad \mu\int_{\mathbb{R}^d} \int_{\mathcal{V}}\int_0^1\left(\int_{\mathcal{V}}T[\mathcal{S},\mathcal{S}^\dagger](\bnu';y) \, \phi(\x,\bnu',y) \, {\rm d}\bnu' -\phi(\x,\bnu,y)\right) \, f(t,\x,\bnu,y) \, {\rm d}y \, {\rm d}\bnu \, {\rm d}\x \\
&\qquad\quad + \lambda\int_{\mathbb{R}^d} \int_{\mathcal{V}}\int_0^1\left(\int_{0}^1 K[\mathcal{S}^\ddagger](y'|y) \, \phi(\x,\bnu,y') \, {\rm d}y' -\phi(\x,\bnu,y)\right) \, f(t,\x,\bnu,y) \, {\rm d}y \, {\rm d}\bnu \, {\rm d}\x \, .
\end{aligned}
$$
}
Then, using the fact that
\begin{eqnarray*}
&& \int_{\mathcal{V}}\int_0^1 \int_{\mathcal{V}}T[\mathcal{S},\mathcal{S}^\dagger](\bnu';y) \, \phi(\x,\bnu',y) \, {\rm d}\bnu' \, f(t,\x,\bnu,y) \, {\rm d}y \, {\rm d}\bnu = \\
&& \qquad \qquad \int_{\mathcal{V}}\int_0^1 \int_{\mathcal{V}}T[\mathcal{S},\mathcal{S}^\dagger](\bnu;y) \, \phi(\x,\bnu,y) \, {\rm d}\bnu \, f(t,\x,\bnu',y) \, {\rm d}y \, {\rm d}\bnu'
\end{eqnarray*}
and
\begin{eqnarray*}
&& \int_{\mathcal{V}}\int_0^1 \int_{0}^1 K[\mathcal{S}^\ddagger](y'|y) \, \phi(\x,\bnu,y') \, {\rm d}y' \, f(t,\x,\bnu,y) \, {\rm d}y \, {\rm d}\bnu = \\
&& \qquad \qquad \int_{\mathcal{V}}\int_0^1 \int_{0}^1 K[\mathcal{S}^\ddagger](y|y') \, \phi(\x,\bnu,y) \, {\rm d}y \, f(t,\x,\bnu,y') \, {\rm d}y' \, {\rm d}\bnu
\end{eqnarray*}
along with definition~\eqref{def:n} of the number density of cells in the phenotypic state $y$, $\rho(t,\x) n(t,\x,y)$, and the notation $\vb = v \hv$, choosing $\phi(\x,\bnu,y) := \xi(\x) \varphi(\bnu,y)$, where $\xi(\x)$ and 
$\varphi(\bnu,y)$ are test functions, we obtain a weak formulation (in the physical space) of the following weak form of the phenotype-structured kinetic equation~\eqref{eq:Boltz_coll_2:strong} for the distribution function $f(t,\x,\bnu,y)$
{\small
\begin{equation}\label{eq:Boltz_coll_2}
\begin{aligned}[b]
\begin{cases}
& \displaystyle{\int_{\mathcal{V}}\int_0^1\varphi(\bnu,y) \, \partial_t f \, {\rm d}y \, {\rm d}\bnu + \nabla_\x \cdot \int_{\mathcal{V}}\int_0^1 \varphi(\bnu,y) \, \vb \, f \, {\rm d}y \, {\rm d}\bnu =}\\
&\qquad\quad \displaystyle{\mu \int_{\mathcal{V}}\int_0^1\varphi(\bnu,y) \, \Big(\rho \, T[\mathcal{S},\mathcal{S}^\dagger](\bnu;y) \, n - f \Big) \, {\rm d}y \, {\rm d}\bnu} \\
&\qquad\quad + \, \displaystyle{\lambda \int_{\mathcal{V}}\int_0^1 \varphi(\bnu,y) \, \left(\int_{0}^1 K[\mathcal{S}^\ddagger](y|y') \, f(t,\x,\bnu,y') {\rm d}y' \, - f\right) {\rm d}y \, {\rm d}\bnu} \, ,
\\\\
& \displaystyle{n(t,\x,y) :=\dfrac{1}{\rho(t,\x)} \int_{\mathcal{V}} f(t,\x,\bnu,y) \, {\rm d}\bnu} \,,
\end{cases}
\end{aligned}
\end{equation}
}
which in the strong form is~\eqref{eq:Boltz_coll_2:strong}.

\section{Additional considerations on macroscopic models}\label{sec:macroclosure}
From the weak form~\eqref{eq:Boltz_coll_2} of the phenotype-structured kinetic equation for $f$, under suitable choices of the test function $\varphi(\bnu,y)$, one can formally derive the governing equations for the moments of $f$, which provide a macroscopic counterpart of the underlying microscopic model. In summary:
\begin{itemize}
\item choosing $\varphi(\bnu,y) := {\bf 1}_{\mathcal{V}}(\bnu) \, \zeta(y)$, where ${\bf 1}_{(\cdot)}$ is the indicator function of the set $(\cdot)$ and $\zeta(y)$ is a test function, we obtain a weak formulation of the following governing equation for the number density of cells in the phenotypic state $y$, which is defined via~\eqref{def:n}, 
\begin{equation}\label{eq:goveqn}
\partial_t \left( \rho n\right) + \nabla_\x \cdot \left(\rho n \, {\bf u}\right) = \lambda \rho \left(\int_{0}^1 K[\mathcal{S}^\ddagger](y|y') \, n(t,\x,y') \, {\rm d}y' \, - n \right) \, ,
\end{equation}
where ${\bf u}$ is the mean velocity of cells in the phenotypic state $y$, which is defined via~\eqref{def:Uy};

\item choosing $\varphi(\bnu,y) :=  \vb \, \zeta(y)$, with $\vb = v \, \hv$, we obtain a weak formulation of the following governing equation for the momentum of cells in the phenotypic state $y$, which is defined via~\eqref{def:nUy}, 
\begin{eqnarray}\label{eq:goveqnUy} 
&&\partial_t \left(\rho n \, {\bf u} \right) + \nabla_\x \cdot \left(\rho n \, {\bf u} \otimes {\bf u} + \rho n \, \mathbb{d}\right) = \mu \rho \left(n \, \ub_T  \, - n \, \ub \right) \nonumber \\
&& \qquad \qquad \qquad + \lambda \rho \, \left(\int_{0}^1 K[\mathcal{S}^\ddagger](y|y') \, n(t,\x,y') \, \ub(t,\x,y') \, {\rm d}y' - n \, \ub\right) \, ,
\end{eqnarray}
where $\ub_T$ is defined via~\eqref{ass:T1} and $\rho n \, \mathbb{d}$ is the second order tensor of cells in the phenotypic state $y$, which is defined via~\eqref{def:nD};

\item choosing $\varphi(\bnu,y) := {\bf 1}_{\mathcal{V}}(\bnu) \, {\bf 1}_{[0,1]}(y)$, we obtain the following governing equation for the cell number density, which is defined via~\eqref{def:m},
\begin{equation}\label{eq:goveqrhoAPP}
\partial_t \rho + \nabla_\x \cdot \left(\rho \, {\bf U}\right) = 0 \, ,
\end{equation}
where ${\bf U}$ is the mean velocity of the cell population, which is defined via~\eqref{def:U};

\item choosing $\varphi(\bnu,y) := \vb \, {\bf 1}_{[0,1]}(y)$, we obtain the following governing equation for the momentum corresponding to the mean velocity of the cell population, which is defined via~\eqref{def:rhoU},
\begin{equation}\label{eq:goveqrhoU}
\partial_t \left(\rho \, {\bf U} \right) + \nabla_\x \cdot \left(\rho \, {\bf U} \otimes {\bf U} + \rho \, \mathbb{D}\right) = \mu \left(\rho \, \Ub_T - \rho \, \Ub\right) \, ,
\end{equation}
where
\begin{equation}\label{def:UTK}
\Ub_T \equiv \Ub_T(t,\x) := \int_{0}^1 \ub_T[\mathcal{S},\mathcal{S}^\dagger](y) \, n(t,\x,y) \, {\rm d}y \, ,
\end{equation}
with $\ub_T$ defined via~\eqref{ass:T1}, and $\rho \, \mathbb{D}$ is the second order tensor of the cell population, which is defined via~\eqref{def:rhoD}.
\end{itemize}

In order to obtain a closed macroscopic model, we can consider appropriately rescaled versions of the system~\eqref{eq:goveqn}-\eqref{eq:goveqrhoU} corresponding to different biological scenarios. We first focus on biological scenarios where phenotypic changes and cell reorientation occur on different time scales. Under these scenarios, introducing a small parameter $\varepsilon \in \mathbb{R}^+$, we consider the case where phenotypic changes occur more frequently than cell reorientation by assuming
\begin{equation}\label{ass:epsscal1}
\lambda = \dfrac{1}{\varepsilon} \, , \quad \mu = O(1) \;\; \text{for } \varepsilon \to 0^+ \, ,
\end{equation}
and the opposite case, i.e. we alternatively assume
\begin{equation}\label{ass:epsscal2}
\mu = \dfrac{1}{\varepsilon} \, , \quad \lambda = O(1) \;\; \text{for } \varepsilon \to 0^+ \, .
\end{equation}
Then we consider biological scenarios where phenotypic changes and cell reorientation occur on similar time scales, which are faster than the time scale of collective spatial dynamics of the cells, i.e. we assume 
\begin{equation}\label{ass:epsscal3APP}
\lambda = \dfrac{1}{\varepsilon} \, , \quad \mu = \dfrac{1}{\varepsilon} \, .
\end{equation}

Throughout this section, we denote by $f$ the limit of $f_{\varepsilon}$ (i.e. the solution to the phenotype-structured kinetic equation~\eqref{eq:Boltz_coll_2:strong} under one of the scalings~\eqref{ass:epsscal1}-\eqref{ass:epsscal3APP}) as $\varepsilon \to 0^+$, with the corresponding macroscopic quantities being defined via~\eqref{def:m}-\eqref{def:rhoD} and~\eqref{def:n}-\eqref{def:nD}. Moreover, we refer the reader to~\eqref{rem:g1} for the definition of $g_1[\mathcal{S}^\ddagger](y)$.

\paragraph{Closure of the macroscopic model under the scaling~\eqref{ass:epsscal1}} Under the scaling~\eqref{ass:epsscal1}, in the asymptotic regime $\varepsilon \to 0^+$:
\begin{itemize}
\item from~\eqref{eq:goveqn} one formally finds
\begin{equation}\label{def:neps0}
n(t,\x,y) = g_1[\mathcal{S}^\ddagger](y) \, ;
\end{equation}
\item from~\eqref{eq:goveqrhoAPP}-\eqref{def:UTK}, using the expression for $n$ given by~\eqref{def:neps0}, one formally obtains the following system
\begin{equation}\label{eq:govern}
\begin{cases}
\partial_t \rho + \nabla_\x \cdot \left(\rho \, {\bf U}\right) = 0 \, ,
\\\\
\partial_t \left(\rho \, {\bf U} \right) + \nabla_\x \cdot \left(\rho \, {\bf U} \otimes {\bf U} + \rho \, \mathbb{D}\right) = \mu \left(\rho \, \Ub_T - \rho \, \Ub\right) \, ,
\end{cases}
\end{equation}
where
\begin{equation}\label{def:UTKeps0}
\Ub_T \equiv \Ub_T[\mathcal{S},\mathcal{S}^\dagger,\mathcal{S}^\ddagger] := \int_{0}^1 \ub_T[\mathcal{S},\mathcal{S}^\dagger](y) \,  g_1[\mathcal{S}^\ddagger](y) \, {\rm d}y 
\end{equation}
with $\ub_T$ being defined via~\eqref{ass:T1}.
\end{itemize}
In order to close the system~\eqref{eq:govern}, we need to find a closed-form expression for $\mathbb{D}$. To this end, building on a moment closure method that is commonly used for transport models of cell migration -- see, for instance, the review~\citep{Hillen2013} -- we make the ansatz
\begin{equation}\label{eq:pclosure1}
p(t,\x,\bnu) = \int_0^1 T[\mathcal{S},\mathcal{S}^\dagger](\bnu;y) \, g_1[\mathcal{S}^\ddagger](y) \, {\rm d}y \, ,
\end{equation}
which follows from assuming that the normalised distribution of the cell population in the phase space, $p$ is fully determined by the environmental conditions. Substituting this ansatz into~\eqref{def:rhoD} gives the following closed-form expression for the variance-covariance matrix
\begin{equation}\label{eq:DEepsscal1}
\mathbb{D} \equiv \mathbb{D}[\mathcal{S},\mathcal{S}^\dagger,\mathcal{S}^\ddagger] :=  \int_{\mathcal{V}} (\vb-{\bf U})\otimes (\vb-{\bf U})  \, \int_0^1 T[\mathcal{S},\mathcal{S}^\dagger](\bnu;y) \, g_1[\mathcal{S}^\ddagger](y) \, {\rm d}y  \,  {\rm d}\bnu\,, 
\end{equation}
which makes it possible to close system~\eqref{eq:govern}.

\paragraph{Closure of the macroscopic model under the scaling~\eqref{ass:epsscal2}} Under the scaling~\eqref{ass:epsscal2}, in the asymptotic regime $\varepsilon \to 0^+$:
\begin{itemize}
\item from~\eqref{eq:goveqnUy} and~\eqref{eq:goveqrhoU} one formally finds, respectively, 
\begin{equation}\label{def:ueps0}
{\bf u}(t,\x,y) = {\bf u}_{T}[\mathcal{S},\mathcal{S}^\dagger](y)
\end{equation}
and
\begin{equation}\label{def:Ueps0}
\Ub(t,\x) = \Ub_T(t,\x) \, ,
\end{equation}
where $\ub_T$ and $\Ub_T$ are defined via~\eqref{ass:T1} and~\eqref{def:UTK};
\item from~\eqref{eq:goveqn} and~\eqref{eq:goveqrhoAPP}, using the expressions for ${\bf u}$ and $\Ub$ given by \eqref{def:ueps0} and \eqref{def:Ueps0}, one formally obtains the following closed system
{\small
\begin{equation}\label{eq:govern2}
\begin{cases}
\displaystyle{\partial_t \left( \rho n\right) + \nabla_\x \cdot \left(\rho n \, {\bf u}_T\right) = \lambda \rho \left(\int_{0}^1 K[\mathcal{S}^\ddagger](y|y') \, n(t,\x,y') \, {\rm d}y' \, - n \right) \, ,} 
\\\\
\partial_t \rho + \nabla_\x \cdot \left(\rho \, \Ub_T\right) = 0 \, .
\end{cases}
\end{equation}
}
\end{itemize}

\paragraph{Closure of the macroscopic model under the scaling~\eqref{ass:epsscal3APP}} Under the scaling~\eqref{ass:epsscal3APP}, in the asymptotic regime $\varepsilon \to 0^+$:
\begin{itemize}
\item from~\eqref{eq:goveqn} and~\eqref{eq:goveqrhoU} one formally finds, respectively, \eqref{def:neps0} and \eqref{def:Ueps0}, where $\Ub_T$ is defined via~\eqref{def:UTKeps0};
\item from~\eqref{eq:goveqrhoAPP}, using the expression for $\Ub$ given by~\eqref{def:Ueps0} and~\eqref{def:UTKeps0}, one formally obtains 
\begin{equation}\label{eq:govern3APP}
\partial_t \rho + \nabla_\x \cdot \left(\rho \, \Ub_T\right) = 0 \, ,
\end{equation}
with $\Ub_T$ being defined via~\eqref{def:UTKeps0}.
\end{itemize}

\begin{remark}
Note that, under the scaling~\eqref{ass:epsscal3APP}, in the asymptotic regime $\varepsilon \to 0^+$ from~\eqref{eq:Boltz_coll_2} one formally finds the expression for $p$ given by~\eqref{eq:pclosure1} -- i.e. if assumptions~\eqref{ass:epsscal3APP} hold then~\eqref{eq:pclosure1} is the actual expression for $p$ and not only a closure approximation -- which makes it possible to compute explicitly all moments of $p$, including the average velocity and the variance-covariance matrix defined via~\eqref{def:U} and~\eqref{def:rhoD}. Moreover, note that, under the scaling~\eqref{ass:epsscal2}, making the ansatz~\eqref{def:neps0} -- i.e. assuming that the normalised distribution of the cell population in the phenotype space $n$ is fully determined by the local environmental conditions -- one could formally reduce the model~\eqref{eq:govern2} to the model~\eqref{eq:govern3APP}.
\end{remark}

\paragraph{Hyperbolic limit of the phenotype-structured kinetic equation} Under the scaling~\eqref{ass:epsscal3APP}, it is also possible to derive a more accurate closed macroscopic model through the hyperbolic limit of the phenotype-structured kinetic equation~\eqref{eq:Boltz_coll_2:strong}. This model consists of a first order in $\varepsilon$ correction to the PDE~\eqref{eq:govern3APP} (i.e. the PDE~\eqref{eq:govern3}), which is derived by means of an asymptotic procedure whereby one starts from the phenotype-structured kinetic equation~\eqref{eq:Boltz_coll_2:strong} under the scaling~\eqref{ass:epsscal3APP} and makes the Chapman-Enskog expansion~\eqref{ass.CE} for $f_{\varepsilon}$ -- see Section~\ref{mesomacromodels}.

\section{Considerations about the scaling~\eqref{ass:epsscal3}}\label{sec:justscal}
Denoting by $t_0$ and $L$ some characteristic time and length scales of the system and by $V$ and $\bar{\rho}$ some reference values of the cell speed and the cell number density, letting
$$
t \rightarrow \dfrac{t}{t_0} \,, \quad {\bf x} \rightarrow \dfrac{{\bf x}}{L} \,, \quad {\bf v} \rightarrow \dfrac{{\bf v}}{V} \,, \quad  \rho \rightarrow \dfrac{\rho}{\bar{\rho}} \,, \quad f \rightarrow \dfrac{f}{\bar{\rho}/V^d}
$$
and
$$
T[\mathcal{S},\mathcal{S}^\dagger](\bnu;y)\rightarrow \dfrac{T[\mathcal{S},\mathcal{S}^\dagger](\bnu;y)}{V^d} \, ,
$$
we obtain the following non-dimensionalised form of the phenotype-structured kinetic equation~\eqref{eq:Boltz_coll_2:strong} for the distribution $f(t,\x,\bnu,y)$
\begin{eqnarray}\label{eq:cinetique_nondim}
\textrm{St} \,  \partial_t f + \vb \cdot \nabla_\x f &=& \dfrac{1}{\textrm{Kn}_{\bnu}}  \left(\rho \, T[\mathcal{S},\mathcal{S}^\dagger](\bnu;y) \, n - f \right) \nonumber \\
&& + \dfrac{1}{\textrm{Kn}_y}  \left(\int_{0}^1 K[\mathcal{S}^\ddagger](y|y') \, f(t,\x,\bnu,y') \,  {\rm d}y' \, - f \right) \, ,
\end{eqnarray}
where the Strouhal number, $\textrm{St}$, and the Knudsen numbers, $\textrm{Kn}_{\bnu}$ and $\textrm{Kn}_y$, are defined as
\begin{equation*}
    \textrm{St}:=\dfrac{L}{V t_0}, \qquad  
    \textrm{Kn}_{\bnu}:=\dfrac{V}{L \mu}, \qquad \textrm{Kn}_y:=\dfrac{V}{L \lambda} \, .
\end{equation*}
Consistently with the experiments corresponding to Figure 1A in~\citep{goodman1989e8}, we choose $t_0=48$h, $V=0.5$cm/h, $L=10$cm, and $\mu=0.5$h$^{-1}$, and we also consider phenotypic changes and spatial movement of cells to occur on similar time scales (i.e. $\lambda \approx \mu$). Hence
\[
\dfrac{L}{V t_0} \approx 1 \, , \qquad \dfrac{V}{L \mu} =10^{-1} \, , \qquad \dfrac{V}{L \mu} \approx \dfrac{V}{L \lambda} \, .
\] 
It is then natural to introduce a small parameter $\varepsilon \in \mathbb{R}^+$ and set 
$$
\textrm{St} = 1 \, , \qquad \textrm{Kn}_{\bnu} = \varepsilon \, , \qquad \textrm{Kn}_{y} = \varepsilon \, .
$$
Substituting into~\eqref{eq:cinetique_nondim} yields 
$$
\partial_t f + \vb \cdot \nabla_\x f = \dfrac{1}{\varepsilon}  \left(\rho \, T[\mathcal{S},\mathcal{S}^\dagger](\bnu;y) \, n - f \right)+ \dfrac{1}{\varepsilon}  \left(\int_{0}^1 K[\mathcal{S}^\ddagger](y|y') \, f(t,\x,\bnu,y') \,  {\rm d}y' \, - f \right) \, ,
$$
which corresponds to~\eqref{eq:Boltz_coll_2:strong} under the scaling~\eqref{ass:epsscal3}, thus indicating that, in the light of the experiments of~\citep{goodman1989e8}, it is reasonable to consider the scaling~\eqref{ass:epsscal3}.

\section{Computation of the correction}\label{sec:SM.fbot}
In this section we report some details about the computation of the correction term derived in ~\eqref{sec:correction}.
First of all, substituting the expressions of $\rho n^{\bot}$ and $\rho p^{\bot}$ into~\eqref{eq:fbot}, one obtains
{\small
\[
\begin{aligned}[b]
f^{\bot}&= \dfrac{1}{2}\left[\nabla_\x \cdot \left(\Ub_{\mathcal{T}} \rho\right) \mathcal{T}-\vb \cdot \nabla_\x \left(\rho \mathcal{T}\right)\right] \\
&\phantom{=} +\dfrac{1}{2} \left[ \nabla_\x \cdot (\rho\Ub_{\mathcal{T}})T_K-\vb \cdot \nabla_\x (\rho T_K)\right] K\\
& \phantom{=}+\dfrac{1}{2}\left[\int_0^1 \nabla_\x\cdot (\rho\Ub_{\mathcal{T}}) K(y) T(\bnu;y) \, {\rm d}y-\int_0^1\nabla_\x \cdot \left[\dfrac{\rho}{2} \, \left({\bf u}_T(y) + {\bf U}_{\mathcal{T}} \right) K (y)\right]T(\bnu;y) \, {\rm d} y\right]K\\
&\phantom{=} +\dfrac{1}{2} \left[ \nabla_\x \cdot (\rho\Ub_{\mathcal{T}})K-\nabla_\x \cdot \left(\dfrac{\rho}{2} \, \left({\bf u}_T + {\bf U}_{\mathcal{T}} \right) K \right)\right] T.
\end{aligned}
\]
}
Differentiating and using the chain rule one finds
\[
-\dfrac{1}{2}\vb \cdot \nabla_\x (\rho (T_K \pm T)) K= -\vb \cdot \nabla_\x(\rho \mathcal{T})+\dfrac{1}{2} \vb \rho (T_K+T)\cdot \nabla_\x K+\dfrac{1}{2}\vb\cdot \nabla_\x (\rho T) K,
\]
and then, rearranging terms and using the chain rule in the first line of $f^{\bot}$, one obtains~\eqref{eq:fbot_det}, which is
{\small
\begin{equation*}
\begin{aligned}[b]
f^{\bot}&= -\left[\mathcal{T}(\vb-\Ub_{\mathcal{T}})\cdot \nabla_\x\rho+ \left(\vb\cdot \nabla_\x \mathcal{T}-\mathcal{T}\nabla_\x\cdot \Ub_{\mathcal{T}}\right)\rho\right]\\
&\phantom{=}+\dfrac{1}{2}\left[\nabla_\x \cdot \left(\rho\Ub_{\mathcal{T}} \right) \mathcal{T}-\vb \cdot \nabla_\x \left(\rho \mathcal{T}\right)\right]\\
&\phantom{=}+ \dfrac{1}{2}\left[\vb \rho (T_K+T)\cdot \nabla_\x K +\vb \cdot \nabla_\x(\rho T)K -\nabla_\x \cdot \left(\dfrac{\rho}{2} \, \left({\bf u}_T + {\bf U}_{\mathcal{T}} \right) K \right) T\right]\\
& \phantom{=}+\dfrac{1}{2}\left[\int_0^1 \nabla_\x\cdot (\rho\Ub_{\mathcal{T}}) K(y) T(\bnu;y) \, {\rm d}y-\int_0^1\nabla_\x \cdot \left[\dfrac{\rho}{2} \, \left({\bf u}_T(y) + {\bf U}_{\mathcal{T}} \right) K(y)\right]T(\bnu;y) \, {\rm d} y\right]K.
\end{aligned}
\end{equation*} 
}
The kinetic correction $f^\bot$ is then needed in order to compute the correction in the macroscopic equation~\eqref{eq:govern3_corr_0}. Specifically, we must compute $\int_0^1\int_{\mathcal{V}} f^\bot \, {\rm d} \bnu \, {\rm d} y$. 
The first line gives rise to the term (see e.g.~\citep{hillen2006m})
\[
-  \, \rho \Ub_{\mathcal{T}}\nabla_\x \cdot \Ub_{\mathcal{T}} - \nabla_\x\cdot(\mathbb{D}_{\mathcal{T}} \rho).
\]
Then, noticing that
\begin{equation*}
\begin{split}
\dfrac{\rho}{2}(T_K+T) \vb \cdot \nabla_\x &K +\dfrac{1}{2}\vb \cdot \nabla_\x (\rho T) K\\
 &= \nabla_\x \cdot \left(\rho \dfrac{T_K+T}{2}\vb K\right) -K \nabla_\x \cdot \left(\vb \dfrac{\rho}{2}(T_K+T) \right) +\dfrac{1}{2} \vb \cdot \nabla_\x (\rho T) K\\
& =\vb \cdot \nabla_\x  (\rho \mathcal{T}) -\dfrac{K}{2}\vb \cdot\nabla_\x  ( \rho T_K),
\end{split}
\end{equation*}
we have 
\begin{equation*}
\begin{split}
&\dfrac{1}{2}\int_{\mathcal{V}}\int_0^1\vb\Big[\vb \cdot \Big(-\nabla_\x (\rho \mathcal{T}(\bnu,y))\\
&\phantom{\dfrac{1}{2}\int_{\mathcal{V}}\int_0^1\vb[\vb \cdot (-\nabla_\x (\rho \mathcal{T}}+\rho(T_K(\bnu)+T(\bnu;y)) \nabla_\x K(y) +\nabla_\x (\rho T(\bnu;y)) K(y)\Big)\Big] \, {\rm d} \bnu \, {\rm d} y\\
&=\dfrac{1}{2}\int_0^1 \int_{\mathcal{V}}\vb \left[\vb \cdot \nabla_\x  (\rho \mathcal{T}(\bnu,y)) -K(y)\vb \cdot\nabla_\x  ( \rho T_K(\bnu))\right] {\rm d} \bnu {\rm d} y\\
&=\dfrac{1}{2}\int_0^1 \int_{\mathcal{V}} \left[\vb \otimes \vb \nabla_\x  (\rho \mathcal{T}(\bnu,y))  - \vb \otimes \vb \nabla_\x  ( \rho T_K(\bnu)) K(y)\right] {\rm d} \bnu \, {\rm d} y \\
&= \dfrac{1}{2} \int_{\mathcal{V}} \left[\vb \otimes \vb \nabla_\x  (\rho T_K(\bnu))  - \vb \otimes \vb \nabla_\x  ( \rho T_K(\bnu))\right] {\rm d} \bnu =0
\end{split}
\end{equation*}
thanks to~\eqref{eq:p0} and~\eqref{ass:K0}.
Moreover,
\[
\dfrac{1}{2}\nabla_\x\cdot (\rho\Ub_{\mathcal{T}})\int_0^1 \int_{\mathcal{V}} \vb\left[ \mathcal{T}(\bnu,y) +\int_0^1  K(y') T(\bnu;y') \, {\rm d}y' K(y) \right]\, {\rm d}\bnu \, {\rm d} y = \Ub_{\mathcal{T}} \nabla_\x \cdot (\rho \Ub_{\mathcal{T}})
\]
thanks to~\eqref{ass:K0} and \eqref{def:UTKeps0red}, and
\begin{equation*}
\begin{split}
\dfrac{1}{2}\int_0^1 \int_{\mathcal{V}}&\vb\Big[\nabla_\x \cdot \Big(\rho \, \big({\bf u}_T(y) + {\bf U}_{\mathcal{T}} \big) K(y) \Big) T(\bnu;y)\\
&+\int_0^1\nabla_\x \cdot \Big(\rho \, \big({\bf u}_T(y') + {\bf U}_{\mathcal{T}} \big) K(y')T(\bnu;y') \, {\rm d} y'\Big) \, K(y)\Big] \, {\rm d} \bnu \, {\rm d} y\\
& = \int_0^1 \nabla_\x \cdot  \Big(\rho \, \big({\bf u}_T(y) + {\bf U}_{\mathcal{T}} \big)  \Big)\ub_T(y) K(y)\, {\rm d}y,
\end{split}
\end{equation*}
thanks to~\eqref{ass:K0} and \eqref{ass:T1}.
In conclusion, we obtain
\begin{equation}\label{eq:govern.corr}
\begin{split}
\partial_t \rho + \nabla_\x \cdot \left[\rho \, \Ub_{\mathcal{T}} \left(1 - \varepsilon \, \nabla_\x \cdot \Ub_{\mathcal{T}} \right)\right]= \varepsilon \, \nabla_\x \cdot \nabla_\x\cdot(\mathbb{D}_{\mathcal{T}} \rho) +\varepsilon \nabla_\x \cdot \mathcal{R} \, ,
\end{split}
\end{equation}
where $\Ub_{\mathcal{T}}$ and $\mathbb{D}_{\mathcal{T}}$ are defined via~\eqref{def:UTKeps0red} and~\eqref{def:DT}, while
\[
\mathcal{R}(\x):= \int_0^1 \nabla_\x\cdot \left(\rho\dfrac{\ub_T(y)+\Ub_{\mathcal{T}}}{2}\right) \ub_T(y) K(y) \, {\rm d} y - \Ub_{\mathcal{T}} \nabla_\x \cdot (\rho \Ub_{\mathcal{T}}).
\]
The latter quantity may be rewritten by performing standard computations. In fact,
\begin{equation*}
\begin{split}
&\int_0^1 \nabla_\x\cdot \left(\rho\dfrac{\ub_T(y)+\Ub_{\mathcal{T}}}{2}\right) \ub_T(y) K(y) \, {\rm d} y = \\
&\int_0^1 \nabla_\x\cdot \left(\rho\dfrac{\ub_T(y)}{2}\right) \ub_T(y) K(y) \, {\rm d} y + \int_0^1 \nabla_\x\cdot \left(\rho\dfrac{\Ub_{\mathcal{T}}}{2}\right) \ub_T(y) K(y) \, {\rm d} y =\\ 
&\dfrac{1}{2}\left[\int_0^1\hspace{-0.2cm} \nabla_\x\rho \ub_T(y) \otimes \ub_T(y) K(y) \, {\rm d} y+ \int_0^1 \hspace{-0.2cm}\rho \ub_T(y)\nabla_\x\cdot\ub_T(y) K(y) \, {\rm d} y  +  \Ub_{\mathcal{T}} \nabla_\x \cdot (\rho \Ub_{\mathcal{T}}) \right]=\\[0.1cm]
&\dfrac{1}{2}\left[\mathbb{C}_T\nabla_\x \rho + \rho \boldsymbol{c}_{T}+\Ub_{\mathcal{T}} \nabla_\x \cdot (\rho \Ub_{\mathcal{T}})\right],
\end{split}
\end{equation*}
with
\[
\mathbb{C}_T := \int_0^1 \ub_T(y) \otimes \ub_T(y) K(y) \, {\rm d} y, \qquad \boldsymbol{c}_T:= \int_0^1 \nabla_\x \cdot (\ub_T(y)) \ub_T(y) K(y) \, {\rm d} y.
\]
As a consequence,
\begin{equation*}
\begin{split}
\mathcal{R}(\x) &= \dfrac{1}{2}\mathbb{C}_T\nabla_\x \rho + \dfrac{1}{2}\rho \boldsymbol{c}_{T}+\dfrac{1}{2}\Ub_{\mathcal{T}} \nabla_\x \cdot (\rho \Ub_{\mathcal{T}}) - \Ub_{\mathcal{T}} \nabla_\x \cdot (\rho \Ub_{\mathcal{T}})\\
&=\dfrac{1}{2}\mathbb{C}_T\nabla_\x \rho + \dfrac{1}{2}\rho \boldsymbol{c}_{T}-\dfrac{1}{2}\Ub_{\mathcal{T}} \nabla_\x \cdot (\rho \Ub_{\mathcal{T}}) \\
&=\dfrac{1}{2} \left[ \mathbb{C}_T\nabla_\x \rho +\rho \boldsymbol{c}_{T} - \Ub_{\mathcal{T}} \otimes \Ub_{\mathcal{T}}\nabla_\x\rho -\rho\Ub_{\mathcal{T}}\nabla_\x \cdot \Ub_{\mathcal{T}}\right].
\end{split}
\end{equation*}
This makes it possible rewriting~\eqref{eq:govern.corr} as~\eqref{eq:govern3_corr_0}.

\section{Definitions of the laminin and fibronectin concentrations and the ECM density used in numerical simulations}\label{sec:numdets}
\begin{figure}[h!]
\centering
\includegraphics[width=\textwidth]{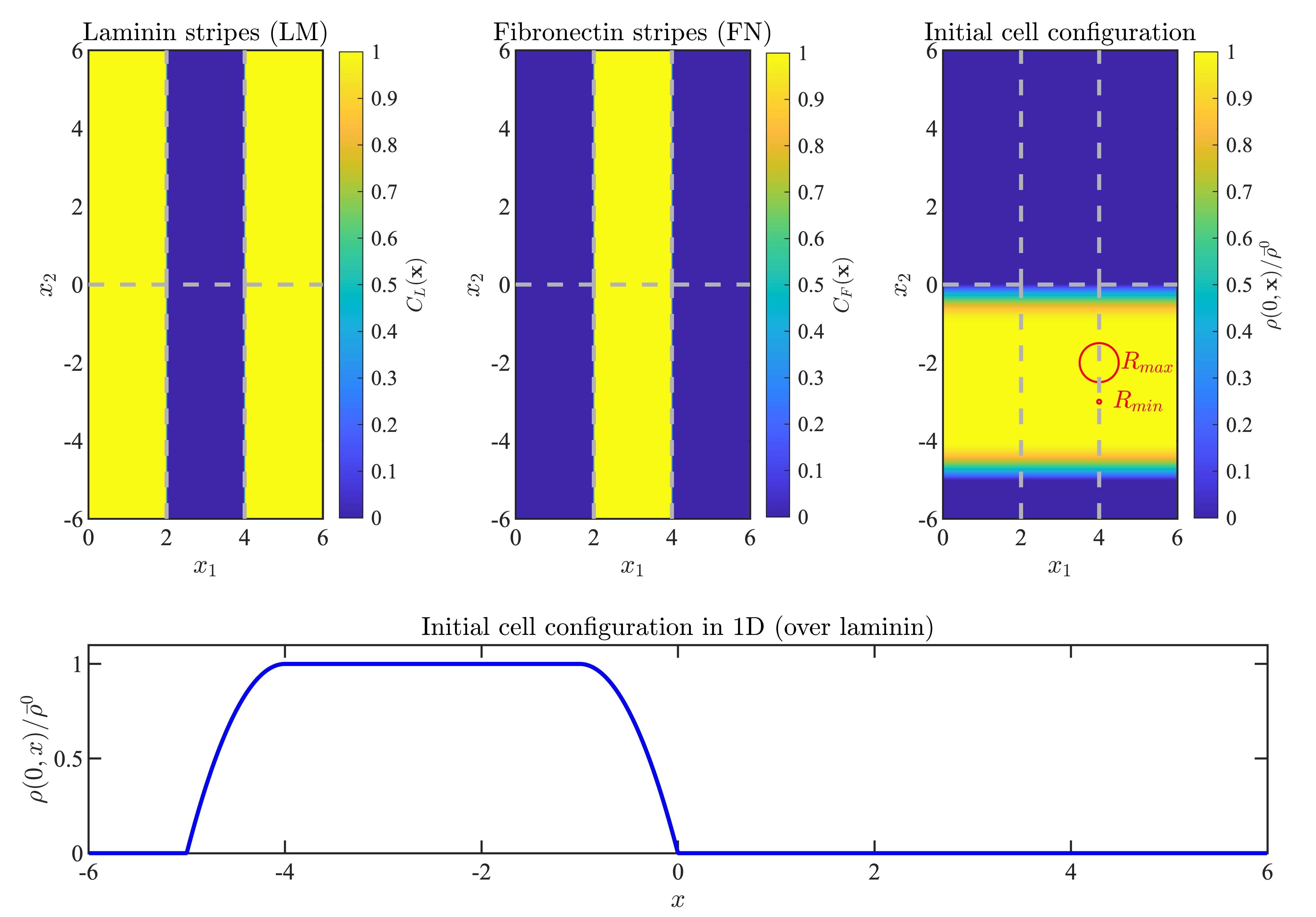}
\caption{\label{fig:ex}Initial conditions for the numerical simulations in the two-dimensional (top row, right panel) and one-dimensional (bottom row) settings, reproducing the experimental set-up of Goodman \textit{et al.}~\citep{goodman1989e8}. In their \textit{in vitro} stripe essay of cell migration, Goodman \textit{et al.} arranged parallel stripes of laminin (cf. top row, left panel) and fibronectin (top row, central panel), and plated cells (myoblasts) at one end of the stripes, evenly distributing them over laminin and fibronectin (top row, right panel).  In this experimental set-up, during cell migration, they observed cells with radius as small as $R_{min}=5\times  10^{-4}$ cm and as large as $R_{max}=5\times  10^{-3}$ cm,  marked in red in the top row, right panel for visual comparison.  In the one-dimensional setting, the cells are assumed to be arranged over a laminin stripe.  The variables $x_1$, $x_2$, and $x$ in the plots are in units of $10^{-2}$cm.  
}
\end{figure}

Consistently with the experimental set-up employed in~\cite[Figure 1A]{goodman1989e8}, illustrated in Figure~\ref{fig:ex}, in the two-dimensional setting, we let laminin and fibronectin be distributed along parallel stripes which run along the $x_2$ direction. Specifically, we use the following definitions, corresponding to the situation where two stripes of laminin are separated by one stripe of fibronectin:
\begin{align}
&{C_L(x_1,x_2) := 
\begin{cases}
\bar{C}_L  & x_1\in \left[0,\frac{1}{3}L_M\right]\cup \left(\frac{2}{3}L_M,L_M\right]\,, \\
0 \,& \text{otherwise}\,,
\end{cases}} 
\quad \forall \,  x_2\in[-L_m,L_M] \, ,
\label{def:L:sim} \\[5pt]
& {C_F(x_1,x_2) := 
\begin{cases}
\bar{C}_F\,& x_1\in \left(\frac{1}{3}L_M,\frac{2}{3}L_M\right] \,,\\
0 \,& \text{otherwise}\,,
\end{cases}} 
\quad \forall \,  x_2 \in [-L_m,L_M] \, ,
\label{def:F:sim} 
\end{align}
with $\bar{C}_L, \bar{C}_F \in \mathbb{R}^+$. Moreover, we assume the ECM density to be uniformly increasing in the $x_2$ direction, and thus use the following definition
\begin{equation}\label{ic:M:sim}
M(x_1,x_2) := M_\text{min} + M_\text{gr} (x_2 +L_m) \quad \forall \,  x_1\in[0,L_M] \, , 
\end{equation}
with $M_\text{min}, M_\text{gr} \in \mathbb{R}^+$.

Analogously, in the one-dimensional setting we use the following definitions 
\begin{equation}\label{def:CLCFM1D}
C_L(x) \equiv \bar{C}_L \, , \quad C_F(x) \equiv 0 \, , \quad M(x) := M_\text{min} + M_\text{gr} (x +L_m) \, .
\end{equation}

Since the laminin and fibronectin concentrations and the ECM density are in non-dimensional form, we choose $\bar{C}_L = 1$, $\bar{C}_F=1$, $M_\text{min}=0.1$, and $M_\text{gr}=1$/cm.

\section{Numerical methods used for solving the PDE~\eqref{eq:govern3_corr}}\label{sec:num:method}
We begin by rewriting the PDE~\eqref{eq:govern3_corr_0} as
\begin{equation}\label{eq:govern3_corr2}
\partial_t \rho + \nabla_\x \cdot \left(\rho \,  \Ub_{\mathcal{T}}^\varepsilon \right)=  \nabla_\x \cdot \left( { \mathbb{D}}_\mathcal{T}^\varepsilon\,\nabla_\x \rho \right) \, ,
\end{equation}
where 
\begin{align}
&\Ub_\mathcal{T}^\varepsilon = \Ub_\mathcal{T}\left( 1 - \varepsilon \nabla_\x \cdot \Ub_\mathcal{T}\right)  - \varepsilon \nabla_\x \cdot {\mathbb{D}}_\mathcal{T} - \frac{\varepsilon}{2}\left( \boldsymbol{c}_T-\Ub_{\mathcal{T}}\nabla_\x \cdot\Ub_{\mathcal{T}}\right)  \, ,\\
&{ \mathbb{D}}_\mathcal{T}^\varepsilon = \varepsilon\, \mathbb{D}_\mathcal{T} + \frac{\varepsilon}{2} \left(\mathbb{C}_T-\Ub_{\mathcal{T}}\otimes\Ub_\mathcal{T}\right)\, , 
\end{align}
and, under the set-up of numerical simulations described in Section~\ref{sec:numsetupmacro}, $\Ub_\mathcal{T}$ and $\mathbb{D}_\mathcal{T}$ are defined via~\eqref{def:UTKeps0red} and~\eqref{def:DT}, $\mathbb{C}_T$ and $\boldsymbol{c}_T$ via~\eqref{def:CT}, with 
$$
\mathcal{S}(t,\x) \equiv M(\x) \, , \quad \mathcal{S}^\dagger(t,\x) \equiv \left(C_L({\bf x}),C_F({\bf x})\right) \, , \quad \mathcal{S}^\ddagger(t,\x) \equiv \left(C_L({\bf x}),C_F({\bf x})\right) \, .
$$
Hence, $\Ub_\mathcal{T}(t,\x) \equiv \Ub_\mathcal{T}(\x)$ and $\mathbb{D}_\mathcal{T}(t,\x) \equiv \mathbb{D}_\mathcal{T}(\x)$, which also imply that $\Ub_\mathcal{T}^\varepsilon(t,\x) \equiv \Ub_\mathcal{T}^\varepsilon(\x)$ and $\mathbb{D}_\mathcal{T}^\varepsilon(t,\x) \equiv \mathbb{D}_\mathcal{T}^\varepsilon(\x)$.

We discretise the spatial domain with a uniform grid of step $\Delta x=0.05$ and solve numerically the PDE~\eqref{eq:govern3_corr2} using an explicit first-order in time mixed finite-difference and finite-volume scheme.  We choose the time-step $\Delta t$ such that the following CFL condition is satisfied
\[
\Delta t \leq \min \left( \frac{\Delta x}{\displaystyle{ \max_{{\bf x}\in\Omega} }\, \left|\Ub_\mathcal{T}^\varepsilon({\bf x})\right| }\,,\,  \frac{(\Delta x)^2}{2d\, \displaystyle{\max_{{\bf x}\in\Omega} }\,\lambda^{loc}_{max}\left( \frac{\mathbb{D}_\mathcal{T}^\varepsilon({\bf x})+ (\mathbb{D}_\mathcal{T}^\varepsilon({\bf x}))^{\mathsf{T}}}{2}\right)}\,  \right)
\]
where $d$ denotes the spatial dimension and $\lambda^{loc}_{max}$ the maximum local eigenvalue of the symmetric part of the corrected anisotropic diffusion coefficient $\mathbb{D}_\mathcal{T}^\varepsilon$, so as to ensure stability of the scheme. 

The numerical approximation of the advection term in~\eqref{eq:govern3_corr2} relies on a finite-volume scheme with a first-order upwind approximation for the advective flux $\rho \,\Ub_\mathcal{T}^\varepsilon$. This is implemented within a MUSCL scheme~\citep{van1979towards} by setting the flux limiter function to zero. Since here $\Ub_\mathcal{T}^\varepsilon$ does not depend on $t$, the scheme includes a correction term obtained from employing the average flux between the current time-step and the next one~\citep{dullemond2008advection}. The numerical approximation of $\Ub_\mathcal{T}^\varepsilon$ relies on a first-order central finite-difference approximation of the first-order spatial derivatives of $\Ub_\mathcal{T}$.

The right-hand side of the PDE~\eqref{eq:govern3_corr2} is treated with first order central finite-difference approximations of the first-order spatial derivatives and linear interpolation of $\rho$ off the grid cell centers. The resulting scheme is second-order in space, and requires a three-point stencil for the approximation at each grid point for simulations in 1D, and a nine-point stencil for the approximation in 2D -- this is due to the presence of second-order mixed derivatives in space, the approximation of which requires the value of $\rho$ at the centres of the diagonal neighbouring grid cells.

The zero-flux boundary conditions~\eqref{eq:cb_macro} are implemented with the use of ghost points. 

This scheme requires the approximation of $\Ub^\varepsilon_\mathcal{T}$ and $\mathbb{D}_\mathcal{T}^\varepsilon$ at the grid cell interfaces, and thus that of $\Ub_\mathcal{T}$, defined via~\eqref{def:UTKeps0red}, $\mathbb{D}_\mathcal{T}$, defined via~\eqref{def:DT}, $\mathbb{C}_T$ and $\boldsymbol{c}_T$, defined via~\eqref{def:CT}, at the same points. For each term, we employ a uniform discretisation of step $\Delta y=0.1$ for the phenotypic domain and a uniform discretisation for the sensing region, which consists of $N\theta=60$ points for $\theta\in[0,\pi]$, and then approximate integrals using a midpoint double Riemann sum. To approximate $\Ub^\varepsilon_\mathcal{T}$ and $\mathbb{D}_\mathcal{T}^\varepsilon$ we must also compute the divergence of $\Ub_\mathcal{T}$ and $\mathbb{D}_\mathcal{T}$ at the grid cell interfaces, which rely on first order central approximations of the spatial derivatives and linear interpolation for $\Ub^\varepsilon_\mathcal{T}$ and $\mathbb{D}_\mathcal{T}^\varepsilon$ off the computed grid.
For consistency with the zero-flux boundary conditions~\eqref{eq:cb_macro}, we replace the points within a sensing region that fall outside the spatial domain $\Omega$ with the nearest boundary points. We remark that preallocating memory for the non-local quantities is crucial for the computational efficiency of the scheme. For further details and numerical optimisation of the computation of non-local terms like those in~\eqref{def:UTKeps0red} and~\eqref{def:DT}, we refer the interested reader to~\citep{gerisch2010approximation}. 

Simulations are performed in {\sc MATLAB}\textsuperscript{\textregistered} and the numerical tests carried out include: mass conservation check; employing different flux limiters in the MUSCL scheme (superbee, Lax-Wendroff, minimod, ospre, Koren, MC, and van Leer); investigating numerical convergence by varying grid step for the spatial variable as well as discretisation step for the phenotypic domain and the sensing region; verifying stability by comparison with the numerical solution obtained using the {\sc MATLAB} function \texttt{ode45}, which is based on an explicit Runge-Kutta method of higher order with time-step dynamically adjusted to control accuracy. 

\bibliographystyle{siam}
\bibliography{TNLC}
\end{document}